\documentclass[conference]{IEEEtran}

\newcommand{\bol}{\boldsymbol}

\newcommand{\ner}{\boldsymbol{r}}

\newcommand{\de}{\,\mathrm{d}}                               
\newcommand{\e}{\operatorname{e}}                               
                     
\newcommand{\inc}{\mathrm{inc}}

\newcommand{\p}{\partial}

\newcommand{\lf}{\left}
\newcommand{\rg}{\right}

\newcommand{\R}{\mathbb{R}}       
\newcommand{\C}{\mathbb{C}}

\newcommand{\mgf}{\bold H}                                        
\newcommand{\elf}{\bold E}

\newcommand{\bnor}{\bold n}

\usepackage{gensymb}
 \usepackage{cite}
\usepackage{multirow}
\usepackage{amsmath}
\usepackage{amsfonts}
\usepackage{amsmath}
\usepackage{amssymb}  
\usepackage{graphicx}
\usepackage{caption}
\usepackage{mathrsfs}
\usepackage{upgreek}
\usepackage{amsthm}
\usepackage{subfig}
\usepackage{booktabs}
\usepackage{authblk}
\usepackage{color}
\usepackage{soul}

\usepackage[T1]{fontenc}

 \usepackage{xcolor}
 \usepackage[normalem]{ulem}

\ifCLASSINFOpdf
\else
\fi
\usepackage{authblk}

\title{Planewave Density Interpolation Methods for the EFIE on Simple and Composite Surfaces}

\author[1]{Carlos~P\'erez-Arancibia}
\author[2]{Catalin Turc}
\author[3]{Luiz~M.~Faria}
\author[4]{Constantine Sideris}
\affil[1]{\small Institute for Mathematical and Computational Engineering, PUC Chile ({\tt cperez@mat.uc.cl})}
\affil[2]{Department of Mathematical Sciences, New Jersey Institute of Technology, USA ({\tt catalin.c.turc@njit.edu})}
\affil[3]{INRIA (Laboratoire POEMS), France ({\tt luiz.faria@ensta-paris.fr})}
\affil[4]{Department of Electrical and Computer Engineering, University of Southern California, USA ({\tt csideris@usc.edu})}

\begin{document}

\maketitle

\begin{abstract}
This paper presents an extension of the recently introduced planewave density interpolation (PWDI) method to the electric field integral equation (EFIE) formulation of problems of scattering and radiation by perfect electric conducting (PEC) objects.  Relying on Kirchhoff integral formula and local interpolation of surface current densities that regularize the kernel singularities, the PWDI method enables off- and on-surface EFIE operators to be re-expressed in terms of integrands that are globally bounded (or even more regular) over the whole domain of integration, regardless of the magnitude of the distance between target and source points. Surface integrals resulting from the application of the method-of-moments (MoM) using Rao-Wilton-Glisson (RWG) basis functions, can then be directly and easily evaluated by means of elementary quadrature rules irrespective of the singularity location. The proposed technique can be applied to simple and composite surfaces comprising two or more simply-connected overlapping components. The use of composite surfaces can significantly simplify the geometric treatment of complex structures, as the PWDI method enables the use of separate non-conformal meshes for the discretization of each of the surface components that make up the composite surface. A variety of examples, including multi-scale and intricate structures, demonstrate the effectiveness of the proposed methodology.
\end{abstract}


\IEEEpeerreviewmaketitle

\section{Introduction}
\IEEEPARstart{A}{s} is well-known, the numerical solution of the classical EFIE by the method of moments (MoM) (also known as the boundary element method (BEM) in other communities) requires numerical evaluation of (weakly) singular integrals, typically defined over planar triangular surface elements~\cite{rao1982electromagnetic}. 
{ Several} analytical, numerical, and hybrid procedures have been developed over the years to numerically evaluate singular integrals in electromagnetic calculations~\cite{tzoulis2005review}. The two most well-established approaches to deal with such integrals { rely} on either the so-called singularity extraction/subtraction technique~\cite{wilton1984potential,graglia1993numerical,caorsi1993theoretical,YlaOijala:2003bn} or the so-called singularity cancellation technique~\cite{Sauter:1996iw,schwab1992numerical,reid2015generalized,taylor2003accurate,polimeridis2010complete}. Singularity extraction/subtraction techniques are based on expressing the integrand as the sum of a simple singular term corresponding to static (Laplace) Green functions, whose integral over triangles can be evaluated in closed form, and a smoother (at least bounded) term whose integral can be directly computed by means of standard quadrature rules. On the other hand, singularity cancellation techniques, such as those based on the Duffy transformation~\cite{duffy1982quadrature} and polar change of variables~\cite{Hackbusch:1994tq}, rely on a certain coordinate transformation that effectively cancels the singularity of the kernel, thus producing a non-singular integral that can be accurately evaluated by means of standard quadrature rules. Recent contributions on this subject include the development of all-analytic techniques~\cite{tihon2018all} that do not rely on numerical integration.

In turn, the off-surface evaluation of the electric field potentials---via which the electromagnetic field is retrieved from the MoM-computed surface current density---involves non-singular integrals that can in principle be computed by means of standard quadrature rules, provided the target point is located sufficiently far from the surface charges. However, as the target point approaches the surface, the kernels become \emph{nearly singular}---a term used to denote functions that possess inordinately large yet not infinite derivatives at a given point. Arguably, a more relevant situation where nearly-singular integrals naturally occur is in scattering problems  involving two or more obstacles that are very close to each other.  Indeed, some of the integrals present in the MoM discretization of the (on-surface) EFIE operator become nearly-singular in this case, as integration needs to be performed on one surface with target points placed on another nearby surface. { Given that this is a time honored integration problem within the boundary integral equation community}, there are also numerous procedures to tackle it. For instance, a certain generalization of the Duffy transformation for nearly-singular integrals was introduced in~\cite{botha2013family,botha2015numerical} and several other techniques in the spirit of the singularity cancellation technique (that can in fact be viewed as some kind of adaptive mesh refinement) can be found in the literature~\cite{khayat2008improved,khayat2005numerical,telles1987self,hayami1994numerical,scuderi2008computation,vipiana2012numerical}.

This paper presents an extension of the PWDI method---put forth in~\cite{plane-wave:2018,perez2019planewave,HDI3D} for the treatment of  weakly singular, hypersingular, and nearly singular integrals arising in boundary integral equations formulations of the Laplace and Helmholtz equations---to the EFIE formulation of problems of scattering/radiation by PEC obstacles represented by simple (either simply or multiply connected) and composite surfaces. Relying on the fact that both on-surface and off-surface electric field potentials can be recast as vector Helmholtz single-layer operators and potentials, the Helmholtz-PWDI method enables both singular and nearly singular EFIE integrals to be expressed in terms of integrands that are bounded or smoother---depending on the density interpolation order---over the whole surface, independent of the distance between target and source points. The resulting PWDI-regularized integrals can then be numerically evaluated by means of simple quadrature rules regardless of the location of Green function singularities. It is worth mentioning that given that the proposed kernel-regularization procedure operates at the continuous level, it can be seamlessly used in conjunction with a variety of low- and high-order discretization schemes available in the literature~\cite{simpson2018isogeometric,cai2001high,cai2002singularity,Bruno2009electromagnetic}, leading to a significantly simpler treatment of singularities, which ultimately reduces the associated implementation effort. Furthermore, unlike other existing singular integration techniques, the PWDI method has the advantage of being universal, in the sense that the same simple regularization procedure is used to deal with singular and nearly singular integrals at the same time. These properties of the PWDI method make it extremely versatile and easy to incorporate in any type of formulation without requiring any specialized approach-dependent techniques. For the sake of preciseness, in this paper, we focus on the classical MoM based on RWG basis functions for representing closed surfaces with planar triangular meshes.

We take advantage of the capability of PWDI to express on- and off-surface electric field potentials in terms of regular integrands, to extend the EFIE formulation to problems involving PEC obstacles modeled as composite surfaces. Unlike simple surfaces, composite surfaces comprise two or more simply-connected overlapping components  corresponding to boundaries of subdomains that make up the whole object. The associated integral representation of the electric field based on the multiple-scattering EFIE formulation, thus involves surface current densities defined on each of the closed-surface components. Coupled integral equations for those currents are obtained by directly enforcing the PEC boundary condition on both the exterior part, corresponding to the actual boundary of the object, and the interior parts where subdomain boundaries overlap. An immediate consequence of this formulation is that it enables the use of non-geometrically conformal meshes (of the closed-surface components) in the MoM discretization of the resulting integral equation system. This property can significantly simplify the geometric treatment of complex PEC structures consisting of several simpler subparts welded together, as it bypasses the need to construct a good-quality single-domain mesh (which is,  in some cases,  a time-consuming and tedious task). This property of the multiple-scattering formulation is also particularly attractive at dealing with problems requiring repeated calculations on a surface that is subject to changes in the position of some of its parts, as is the case in some optimal-design problems~\cite{weile1997genetic,uler1994utilizing,jorgensen2001efficient,koper2004aircraft}. No re-meshing and no re-computation of the diagonal block of the impedance matrix is required in this case, by virtue of the translational invariance of the Green function. This extended multiple-scattering EFIE formulation for composite surfaces shares some similarities with domain decomposition methods based on boundary integral equations~\cite{peng2011integral,peng2015domain,bautista2015nonconformal}. However, the present approach is simpler both conceptually and computationally, as no direct enforcement of the continuity of surface currents is required wherever two or more closed surfaces overlap. In Sec.~VII~D., we demonstrate the advantages of our composite surface approach by using it to analyze a monopole antenna reflector array, where we are able to obtain a significant savings by reusing the mesh and matrix blocks of a single monopole for all of the elements in the array.

This paper is organized as follows. Section~\ref{sec:3d_electro} outlines the EFIE formulation for the solution of problem of electromagnetic scattering by PEC obstacles modeled in terms of simple (simply or multiply connected) surfaces. Section~\ref{sec:MoM} briefly describes the classical MoM discretization of the EFIE. Section~\ref{sec:sef_intersec} then introduces the multiple-scattering EFIE formulation for composite surfaces. Closed-form and purely numerical planewave density interpolation procedures are presented in Section~\ref{sec:PWDI}. Details of the implementation of the proposed methodology are provided in Sec.~\ref{sec:imp_details}. Finally, Sec.~\ref{sec:num_ex} presents a variety of numerical examples.

\section{Problem formulation}\label{sec:3d_electro}
We start off by presenting the problem of time-harmonic electromagnetic scattering of an incident  wave field $(\elf^\inc,\mgf^\inc)$ which illuminates a (possibly multiply connected) bounded PEC object $\Omega\subset\R^3$.  The total electromagnetic field $(\elf,\mgf)$ satisfies the homogeneous Maxwell's equations 
\begin{equation}\label{eq:PDE}
\nabla\times  \elf  -ik \mgf =\bold 0\  \mbox{and}\  \nabla\times  \mgf + ik \elf =\bold 0\ \mbox{in}\ \R^3\setminus\overline{\Omega},
\end{equation}
where $k=2\pi/\lambda=\omega\sqrt{\epsilon\mu}$ is the wavenumber, with $\omega$ denoting the angular frequency,  and  $\epsilon>0$ and $\mu>0$ denoting the electric permittivity and the magnetic permeability of the medium surrounding $\Omega$, respectively. 
Expressing the total field in the form
$\lf(\elf^t,\mgf^t\rg) = \lf(\elf^s,\mgf^s\rg)+ \big(\elf^\inc,\mgf^\inc\big)$
 we obtain that the scattered field $(\elf^s,\mgf^s)$ satisfies Maxwell's equations~\eqref{eq:PDE}
together with the boundary condition
\begin{equation}
\bold n\times \elf^s = -\bold n\times \elf^\inc\ \mbox{on}\ \Gamma
 \label{eq:scattered_maxwell_3D_BC}
\end{equation}
and the Silver-M\"uller radiation condition
\begin{equation}
\lim_{|\ner|\to\infty}|\ner|\left(\nabla\times\elf^s\times\frac{\ner}{|\ner|}-ik\,\elf^s\right)=\bold0,
 \label{eq:scattered_maxwell_3D_RC}
\end{equation}
which holds uniformly in all directions $\ner/|\ner|$~\cite{COLTON:1983}. Per usual, throughout the paper the symbol $\bold n$  denotes the outer unit normal to the surface~$\Gamma$.  (In what follows we focus on the scattering problem only, as the radiation problem is analogous.)

The classical EFIE formulation of the scattering problem is derived from an integral representation whereby the scattered electric field is expressed as
\begin{equation}
\elf^s(\ner) = (\mathcal E\bold J)(\ner):= ik\mathcal (\mathcal E_1\bold J)(\ner)-\frac{1}{ik}(\mathcal E_2\bold J)(\ner) \label{eq:EFIE_pot}
\end{equation}
for $\ner\in\R^3\setminus\overline\Omega$, 
in terms of the integral operators 
\begin{subequations}\begin{align}
&(\mathcal E_1\bold J)(\ner):=\int_{\Gamma}G(\ner,\ner')\bold J(\ner') \de s'\quad\mbox{and}\label{eq:vec_sl}\\
&(\mathcal E_2\bold J)(\ner):=\nabla\int_{\Gamma} G(\ner,\ner')\ \nabla'_{s}\cdot\bold J (\ner') \de s',\label{eq:grad_sl}
\end{align}\label{eq:op_in_vol}\end{subequations}
where $G(\ner,\ner') = (4\pi|\ner-\ner'|)^{-1}\e^{ik|\ner-\ner'|}$ is the free-space Green function of the Helmholtz equation, and where $\Gamma$ denotes, for the time being,  the boundary of $\Omega$. The scattered electromagnetic field thus represented, which is given by  $(\mathcal E\bold J,(ik)^{-1}\nabla\times \mathcal E\bold J)$, is an exact solution of the Maxwell system~\eqref{eq:PDE}. The surface current density $\bold J$, which is a vector field tangential to the surface, is then the unknown we aim to solve for. Enforcing the PEC boundary condition~\eqref{eq:scattered_maxwell_3D_BC}, an integral equation for the unknown surface currents can be found. To this end, the tangential components of~\eqref{eq:EFIE_pot} are directly evaluated on the surface via the relation~\cite{COLTON:1983}
\begin{equation}\label{eq:limit_Epot}
\lim_{\delta\to +0}  \bnor\times (\mathcal E\bold J)(\ner\pm \delta \bnor) = (\mathsf E\bold J)(\ner),\quad\ner\in\Gamma,
\end{equation}
 where $\mathsf E=ik\mathsf E_1-(ik)^{-1}\mathsf E_2$ is the so-called EFIE operator which is given in terms of
\begin{subequations}\begin{align}
&(\mathsf E_1\bold J)(\ner):=\bold n(\ner)\times\int_{\Gamma}G(\ner,\ner')\bold J(\ner') \de s'\quad\mbox{and}\\
&(\mathsf E_2\bold J)(\ner):=\bold n(\ner)\times\nabla\int_{\Gamma} G(\ner,\ner')\ \nabla'_{s}\cdot\bold J (\ner') \de s'.\label{eq:E2}
\end{align}\label{eq:op_in_surf}\end{subequations}
It thus follows from~\eqref{eq:EFIE_pot} and~\eqref{eq:limit_Epot} that the PEC boundary condition~\eqref{eq:scattered_maxwell_3D_BC}   yields the well-known EFIE:
\begin{equation}\label{eq:E_EFIE}
\mathsf E\bold J= -\bnor\times\elf^\inc \ \mbox{on} \ \Gamma,\end{equation}
for the unknown surface current density $\bold J$.  (The mathematical details behind the derivations presented in this section can be found in~\cite{hsiao1997mathematical}.)

\section{Method of moments}\label{sec:MoM}
Throughout this paper, we focus on the classical discretization of the EFIE by means of the MoM using the RWG basis functions~\cite{rao1982electromagnetic}. We thus consider a triangulation of the closed surface $\Gamma$ which is  assumed to be given by the union of~$N_h$ planar triangles $T_j$ (i.e.,~$\Gamma=\bigcup_{j=1}^{N_h}T_j$) with maximum edge size $h>0$.  In order to solve~\eqref{eq:E_EFIE}, the unknown $\bold J$ is expanded~as
\begin{equation}
\bold J(\ner) \approx \sum_{n=1}^N I_n\bold f_n(\ner)\quad (\ner\in\Gamma),\label{eq:current_exp}
\end{equation}
 in terms of the div-conforming RWG basis functions~$\bold f_n$ defined on the $N$  edges of the surface mesh (more details are provided in Sec.~\ref{sec:RWG}).  The expansion coefficients $I_n$, $1\leq n\leq N$, are  obtained by substituting  the approximation~\eqref{eq:current_exp} in the integral equation~\eqref{eq:E_EFIE}, which yields
\begin{equation}\sum_{n=1}^NI_n\mathsf E\bold f_n=-\bnor\times\elf^\inc\quad\mbox{on}\quad \Gamma.\label{eq:galerkin}\end{equation}
Testing~\eqref{eq:galerkin} against the curl-conforming basis functions $\bnor\times \bold f_n$, $1\leq n\leq N$,  the following linear system
\begin{equation}Z\bold I = \bold V\label{eq:lin_sym}\end{equation} is achieved,
where the relevant matrices and vectors are  $(\bold I)_n =I_n$, $(\bold V)_m = -\lf\langle\bnor\times\bold f_m,\bnor\times\elf^\inc\rg\rangle=-\lf\langle\bold f_m,\elf^\inc\rg\rangle$ and $Z=ikZ_1-(ik)^{-1}Z_2$ with $(Z_j)_{m,n} =\lf\langle\bnor\times\bold f_m,\mathsf E_j\bold f_n\rg\rangle$, $j=1,2$.
In detail, the entries of impedance matrix components are 
\begin{subequations}
\begin{align}
(Z_1)_{m,n}
=&\displaystyle\int_{\Gamma}\bold f_m (\ner)\cdot\lf\{\int_{\Gamma}G(\ner,\ner')\bold f_n (\ner') \de s'\rg\}\de s,\label{eq:Z1}\\
(Z_2)_{m,n} =&\displaystyle \int_{\Gamma}\bold f_m(\ner) \cdot 
\lf\{\nabla\int_{\Gamma}  G(\ner,\ner') \ \nabla'_{s}\cdot\bold f_n (\ner') \de s'\rg\}\de s\nonumber.\label{eq:Z2p}
\end{align}
for $1\leq m,n\leq N$. As is well known, resorting to integration by parts---in order to transfer the gradient to the test basis function~$\bold f_m$---the last integral above can be re-expressed as
\begin{equation}
(Z_2)_{m,n}
=-\int_{\Gamma}\nabla_s\cdot\bold f_m(\ner) 
\lf\{\int_{\Gamma}G(\ner,\ner') \ \nabla'_{s}\cdot\bold f_n (\ner') \de s'\rg\}\de s.\label{eq:Z2}
\end{equation}\label{eq:entries}\end{subequations}
We note here that the kernels in~\eqref{eq:entries} exhibit only a weak (integrable) $\mathcal O(|\ner-\ner'|^{-1})$ singularity.

Once the approximate currents have been obtained, the electric and magnetic fields can be retrieved by means of the off-surface operator~\eqref{eq:EFIE_pot}. The kernels present in~\eqref{eq:EFIE_pot} are smooth but may become nearly-singular as the target point $\ner\in\R^3\setminus\overline\Omega$ approaches the boundary~$\Gamma$. A similar phenomenon arises in problems involving two (or more) simply connected domains, say, $\Omega_1$ and $\Omega_2$ with $\Omega=\Omega_1\cup\Omega_2$ and  $\overline\Omega_1\cap\overline\Omega_2=\emptyset$, that are very close to  each other (see, for example, Fig.~\ref{fig:scatt_sphere}). Some of the impedance matrix entries~\eqref{eq:entries} in this case involve outer integrals over $\Gamma_1=\p\Omega_1$ and inner integrals over  $\Gamma_2=\p\Omega_2$, thus leading to nearly-singular behavior of the kernel present in the inner integral as the distance $|\ner-\ner'|$, with $\ner\in\Gamma_1$ and $\ner'\in\Gamma_2$, may become very small but not zero. 

\section{Composite surfaces}\label{sec:sef_intersec}
We have so far dealt with the case  where the PEC object~$\Omega$ consists of a collection of nonintersecting simply connected domains. Assume now that  $\Omega$ is a simply connected domain {which can be represented naturally as the union of several non-overlapping subdomains.} In order to fix the ideas, let us assume for simplicity that $\Omega= \Omega_1\cup\Omega_2$ with $\Omega_1\cap\Omega_2=\emptyset$, but $\Gamma_1\cap\Gamma_2\neq\emptyset$, where $\Gamma_1=\p\Omega_1$ and~$\Gamma_2=\p\Omega_2$, that is, the two domains  $\Omega_1$ and $\Omega_2$ share in common a point, a curve, or an open surface (see, for example, Figs.~\ref{fig:city_meshes} and~\ref{fig:IC}). 

Instead of using the  EFIE formulation posed on the boundary of the obstacle~$\p\Omega$, we prefer to employ an extended multiple-scattering EFIE formulation posed on the composite surface $ \Gamma=\Gamma_1\cup\Gamma_2$ with components $\Gamma_1$ and~$\Gamma_2$. Just as in the simple-surface EFIE formulation presented above, we express the scattered field $\elf^s$ as~\eqref{eq:EFIE_pot} in terms of the off-surface operator $\mathcal E$ integrating now over $\Gamma=\Gamma_1\cup\Gamma_2$ and featuring a current density function  $\bold J=[\bold J_1\ \bold J_2]^\top$ also defined on $\Gamma$. The enforcement of the PEC boundary condition~\eqref{eq:scattered_maxwell_3D_BC} on both $\Gamma_1$ and $\Gamma_2$ yields the multiple-scattering EFIE:
\begin{equation}\label{eq:E_EFIE_comp}
\begin{bmatrix}\mathsf E_{11} & \mathsf{E}_{12}\\ 
\mathsf E_{21} & \mathsf{E}_{22}
\end{bmatrix}
\begin{bmatrix}\bold J_1\\ \bold J_2\end{bmatrix}= -\begin{bmatrix}\bnor_1\times\elf^\inc_1\\  \bnor_2\times\elf^\inc_2\end{bmatrix}\ \mbox{on} \ \Gamma=\Gamma_1\cup\Gamma_2.\end{equation}
In equation~\eqref{eq:E_EFIE_comp}, the electric field operators $\mathsf{E}_{j\ell}$ are defined just as in equations~\eqref{eq:op_in_surf} with $\ner\in\Gamma_j$ and the domain of integration $\Gamma_\ell$ where $\{j,\ell\}=\{1,2\}$. 

The main advantage of the multiple-scattering EFIE formulation~\eqref{eq:E_EFIE_comp} is that it is amenable to MoM discretizations using separate triangular meshes on $\Gamma_1$ and respectively on $\Gamma_2$ that do not conform on $\Gamma_1\cap\Gamma_2$. Their MoM discretization leads to linear systems similar to~\eqref{eq:lin_sym}, whose sub matrices $Z_1$ and $Z_2$ have entries defined in~\eqref{eq:Z1} and~\eqref{eq:Z2}, respectively.  It is worth noting here that the MoM discretization of the EFIE formulation~\eqref{eq:E_EFIE_comp} gives rise to additional integration challenges not present in the classical EFIE formulation. Indeed, the MoM discretization of the composite surface EFIE formulation~\eqref{eq:E_EFIE_comp} requires numerical evaluation of  nearly singular integrals of the form
\begin{equation}\label{eq:comp_surf_for}
\int_{\Gamma_j}  G(\ner,\ner') \bold f_n (\ner') \de s'\ \mbox{and}\  \int_{\Gamma_j} G(\ner,\ner') \ \nabla'_{s}\cdot\bold f_n (\ner') \de s',
\end{equation}
 at target points $\ner\in\Gamma_i\cap\Gamma_j$, $i\neq j$, using meshes on $\Gamma_i$ and $\Gamma_j$  that may not conform to each other on $\Gamma_1\cap\Gamma_2$. We address this as well as other integration issues in the next section, where we show that all the singular and nearly-singular kernels arising in EFIE formulations on simple and composite surfaces can be regularized using the PWDI method~\cite{perez2019planewave}, thus enabling the use of elementary quadrature rules in the practical implementation of the MoM.

\section{Planewave density interpolation}\label{sec:PWDI}
This section is devoted to the presentation of the PWDI method for the regularization of the  kernels associated to the integral operators in~\eqref{eq:op_in_vol} and~\eqref{eq:op_in_surf} discretized using RWG basis functions. 

Throughout this section we make use of the Helmholtz single- and double-layer potentials defined, respectively, as 
\begin{subequations}\begin{align}
(\mathcal  S \varphi)(\ner):=&\int_\Gamma G(\ner,\ner')\varphi(\ner')\de s\quad\mbox{and}\label{eq:sl_pot}\\
(\mathcal D\varphi)(\ner) :=& \int_{\Gamma}\frac{\p G(\ner,\ner')}{\p \bnor'}\varphi(\ner')\de s',\quad \ner\in\R^3\setminus\Gamma,\label{eq:dl_pot}
\end{align}\end{subequations}
 as well as  the associated single- and double-layer operators, defined as
\begin{subequations}\begin{align}
(\mathsf S \varphi)(\ner):=&\int_\Gamma G(\ner,\ner')\varphi(\ner')\de s\quad\mbox{and}\label{eq:sl_op}\\
(\mathsf K\varphi)(\ner) :=& \int_{\Gamma}\frac{\p G(\ner,\ner')}{\p \bnor'}\varphi(\ner')\de s',\quad \ner\in\Gamma,\label{eq:dl_op}
\end{align}\end{subequations}
respectively. We recall here that  potentials and operators are connected by means of the jump relations~\cite{COLTON:1983} 
\begin{subequations}\begin{align}
\lim_{\delta\to 0^+}  (\mathcal S\varphi)(\ner\pm \delta \bnor) &= (\mathsf S\varphi)(\ner)\quad\mbox{and}\label{eq:limit_Spot}\\
\lim_{\delta\to 0^+}  (\mathcal D\varphi)(\ner\pm \delta \bnor) &= \pm \frac{\varphi(\ner)}{2}+(\mathsf K\varphi)(\ner)\quad (\ner\in\Gamma)\label{eq:limit_Dpot},
\end{align}\label{eq:jumps}\end{subequations}
which hold almost everywhere.

The proposed regularization technique relies on two simple observations:
\begin{itemize}
\item The off-surface EFIE operator $\mathcal E$ can be expressed in terms of a vectorial  single-layer potential---referred to as~$\mathcal E_1$ in~\eqref{eq:vec_sl}---and the gradient of a scalar  single-layer potential---referred to as~$\mathcal E_2$ in~\eqref{eq:grad_sl}.
 \item The  MoM discretization of the on-surface EFIE operator~$\mathsf E$ can be expressed in terms of single-layer operators. In fact, in view of~\eqref{eq:entries} it is clear that forming the impedance matrix $Z$ entails evaluation of integrals corresponding to the Galerkin BEM discretization of the Helmholtz single-layer operator.   \end{itemize}

In what follows we thus restrict ourselves to describing the regularization of the single-layer potential~\eqref{eq:sl_pot}, and the regularization of double integrals of the form
\begin{align}
\langle \psi,\mathsf S \varphi\rangle=& \int_{\Gamma} \psi(\ner) \int_{\Gamma} G(\ner,\ner')\varphi(\ner')\de s' \de s\label{eq:single_layer_op}
\end{align}
 where $\psi$ and $\varphi$ are scalar densities, which may correspond to either individual components of the RWG basis functions $\bold f_n$ (in the case of $\mathsf E_1$) or their surface divergence $\nabla_s\cdot \bold f_n$ (in the case of $\mathsf E_2$).
 
The proposed kernel-regularization technique relies on interpolation of the relevant densities by means of linear combinations of planewaves of the form 
\begin{equation}\label{eq:pwdi}
\Phi(\ner',\ner_0) = \sum_{\ell=1}^L c_\ell(\ner_0) \e^{ik\bol d_\ell\cdot(\ner_0-\ner')},
\end{equation} where $\ner'\in\R^3$ and $\ner_0\in\Gamma$, with $\bol d_\ell\in\R^3$, $|\bol d_\ell|=1$, $1\leq \ell\leq L$ denoting planewave directions, which may or may not depend on~$\ner_0\in\Gamma$. Since this linear combination satisfies the homogeneous Helmholtz equation 
$${\nabla'}^2\Phi(\ner',\ner_0)+k^2\Phi(\ner',\ner_0)=0\mbox{ for all }\ner'\in\R^3,$$
it follows from Kirchhoff integral formula~\cite{volakis2012integral} (also known as Green's third identity or extinction theorem) that the single-layer potential~\eqref{eq:sl_pot} can be expressed as~\cite{perez2019planewave}
\begin{equation}\label{eq:reg_SL}\begin{split}
(\mathcal  S \varphi)(\ner)&=\int_\Gamma G(\ner,\ner')\lf\{\varphi(\ner')-\Phi_n(\ner',\ner_0)\rg\}\de s'+\\
&\bol 1_{\Omega}(\ner)\Phi(\ner,\ner_0)+\int_{\Gamma}\frac{\partial G(\ner,\ner')}{\partial\bnor'}\Phi(\ner',\ner_0)\de s'
\end{split}\end{equation}
for all $\ner\in\R^3\setminus\Gamma$. Here, $\Phi_n(\ner',\ner_0) =\bnor'\cdot\nabla'\Phi(\ner',\ner_0)$ and the function~$\bol1_{\Omega}$  denotes the indicator function of the domain $\Omega$; that is, $\bol 1_{\Omega}(\ner)=1$ if $\ner\in\Omega$  and $\bol 1_\Omega(\ner)=$ if $\ner\in\R^3\setminus\overline\Omega$. Note that the last integral above corresponds to the Helmholtz double-layer potential~\eqref{eq:dl_pot} applied to the interpolant $\Phi(\cdot,\ner_0)$.

Therefore, using the jump relations~\eqref{eq:jumps} the following equivalent formula for~\eqref{eq:single_layer_op} is found:
\begin{equation}\label{eq:reg_SLDI}\begin{split}
&\langle\psi,\mathsf S \varphi\rangle=\frac{1}{2}\int_\Gamma\psi(\ner)\Phi(\ner,\ner_0)\de s+\\
&\int_\Gamma \psi(\ner)\int_\Gamma G(\ner,\ner')\lf\{\varphi(\ner')-\Phi_n(\ner',\ner_0)\rg\}\de s'\de s+\\
&\int_\Gamma\psi(\ner)\int_\Gamma\frac{\p G(\ner,\ner')}{\p \bnor'}\Phi(\ner',\ner_0)\de s'\de s.
\end{split}\end{equation}

Note that, up to this point, we have treated $\ner_0\in\Gamma$ as a free parameter that we can choose at our convenience. The key idea underlying the proposed regularization technique is that, in order to achieve bounded or even smoother integrands in~\eqref{eq:reg_SLDI} it suffices to select $\ner_0=\ner$ and  require $\Phi_n(\cdot,\ner)$ and $\Phi(\cdot,\ner)$  to approximate $\varphi$ and the zero density, respectively, at the point $\ner'=\ner$ (precisely where the integral kernels become singular). This is achieved here by asking  $\Phi(\cdot,\ner)$ to satisfy certain pointwise interpolation conditions. In detail, letting $M_1,M_2\geq0$, these conditions are:
\begin{subequations}\begin{align}
\lim_{\ner'\to \ner} {\partial^{\prime}}^{\alpha}_s\Phi(\ner',\ner) =&~0,\ \forall |\alpha|\leq M_1,\mbox{ and}\label{eq:dir_cond}\\
\lim_{\ner'\to \ner} {\partial^{\prime}}^{\alpha}_s\lf\{\varphi(\ner')-\Phi_n(\ner',\ner)\rg\} =&~0,\  \forall  |\alpha|\leq M_2,\label{eq:neu_cond}
\end{align}\label{eq:Iconds}\end{subequations}
where  ${\p'}^\alpha_s$,  with $\alpha=(\alpha_1,\alpha_2)\in \mathbb{N}^2$ and $|\alpha| = \alpha_1+\alpha_2$, denotes the $|\alpha|$-th order  tangential derivative on $\Gamma$ (with respect to $\ner'$). It follows from~\eqref{eq:Iconds} that the integrands in~\eqref{eq:reg_SLDI} satisfy
\begin{subequations}\begin{align}\label{eq:control_ho_kernel}
\lf|G(\ner,\ner')\lf\{\varphi(\ner')-\Phi_n(\ner',\ner)\rg\}\rg|&\lesssim |\ner-\ner'|^{M_2}, \\
\left|\frac{\partial G(\ner,\ner')}{\partial\bnor'}\Phi(\ner',\ner)\right|&\lesssim |\ner-\ner'|^{M_1},
\end{align}\end{subequations}
in a neighborhood of $\ner\in\Gamma$, provided~$\Phi$ fulfills the conditions~\eqref{eq:Iconds}. This means that the integrands in~\eqref{eq:reg_SLDI}  become bounded or even smoother functions of $\ner'\in\Gamma$  whenever $M_1,M_2\geq 0$ and, thus, their integral  can be approximated by standard quadrature rules. 

(Note that in the derivations presented above we have assumed that the scalar density $\varphi$ is sufficiently smooth in a neighborhood $B$ of the interpolation point $\ner\in\Gamma$. Concurrently, we have assumed that there exists a sufficiently smooth local parametrization  $\bold X:D\subset\R^2\to B\subset\Gamma$  around $\ner\in B$ so that the tangential derivatives of a density function $\varphi:\Gamma\to\C$ exist and are given by $\p^\alpha_s\varphi(\ner) = \frac{\p^{\alpha_1}}{\p \xi_1^{\alpha_1}}\frac{\p^{\alpha_2}}{\p \xi_2^{\alpha_2}}\varphi(\bold X(\xi_1,\xi_2))$, with  $\alpha = (\alpha_1,\alpha_2)$.  In general, none of these assumptions hold globally. The scalar components of the RWG basis functions, for example, are not even continuous on the whole surface $\Gamma$. Therefore, special care has to be taken in selecting quadrature points for the evaluation of the outer integrals in~\eqref{eq:reg_SLDI}, to make sure that these assumptions are actually satisfied at those points $\ner\in\Gamma$. See~Section~\ref{sec:quad_rule} for  details.)

Similarly, the regularization of the nearly-singular kernels in the on-surface integral operators~$\mathcal E$ in~\eqref{eq:EFIE_pot}---at points $\ner\in\R^3\setminus\Gamma$ close to but not on~$\Gamma$---uses the fact that $\mathcal E_1$ (resp.~$\mathcal E_2$) can be expressed in terms of the Helmholtz single-layer potential (resp. gradient of the single-layer potential) applied to a vector (resp.~scalar) density function. Regularization of the kernels in this case, can be effected by selecting $\ner_0\in\Gamma$ in~\eqref{eq:reg_SL} as the projection of the target point $\ner\in\R^3\setminus\Gamma$ on the surface, i.e., $\ner_0={\rm argmin}_{r'\in\Gamma}|\ner-\ner'|$~\cite{HDI3D,perez2019planewave}.  Doing so, the integrands in the single-layer potentials present in $\mathcal E_1$ satisfy 
\begin{subequations}\label{eq:control_ho_kernel_pot}\begin{align}
\lf|G(\ner,\ner')\lf\{\varphi(\ner')-\Phi_n(\ner',\ner_0)\rg\}\rg|&\lesssim |\ner_0-\ner'|^{M_2}, \\
\left|\frac{\partial G(\ner,\ner')}{\partial\bnor'}\Phi(\ner',\ner_0)\right|&\lesssim |\ner_0-\ner'|^{M_1-1},
\end{align}\end{subequations}
while the kernels in the gradient of the single-layer potential present in $\mathcal E_2$ satisfy
\begin{subequations}\label{eq:control_ho_grad_kernel_pot}\begin{align}
\lf|\nabla G(\ner,\ner')\lf\{\varphi(\ner')-\Phi_n(\ner',\ner_0)\rg\}\rg|&\lesssim |\ner_0-\ner'|^{M_2-1}, \\
\left|\nabla\frac{\partial G(\ner,\ner')}{\partial\bnor'}\Phi(\ner',\ner_0)\right|&\lesssim |\ner_0-\ner'|^{M_1-2}.
\end{align}\end{subequations}

Exactly the same strategy can be applied to the regularization of the kernels in  integrals of the form~\eqref{eq:comp_surf_for}  arising in the multiple-scattering EFIE formulation. The optimal point~$\ner_0$ in the corresponding regularized form of the single-layer potential  is $\ner_0={\rm argmin}_{\ner'\in\Gamma_j}|\ner-\ner'|$ where $\ner\in\Gamma_i$ in this case.  (The actual practical procedure used to select the regularization point $\ner_0$ used in the MoM discretization of the EFIE is discussed in Section~\ref{sec:quad_rule}.)

The next two sections present  procedures to construct planewave density interpolants~\eqref{eq:pwdi}: a low-order ($M_1=M_2=1$) analytical procedure, and a higher-order least squares procedure.

\subsection{Closed-form PWDI}\label{sec:analy_app}
In order to construct the planewave density interpolant~\eqref{eq:pwdi} we rewrite it as
\begin{equation}
\Phi(\ner',\ner) :=\sum_{|\alpha|=0,1}\p^\alpha_s\varphi(\ner)\Phi_\alpha(\ner',\ner),\label{eq:interpolants}
\end{equation}
where each function  $\Phi_\alpha$ for $|\alpha|\leq 1$ is a linear combinations of planewaves. Therefore, according to the interpolation conditions~\eqref{eq:Iconds} for $M_1=M_2=1$, the expansion functions in~\eqref{eq:interpolants} must fulfill
\begin{equation}\label{eq:point_conditions}
\p^{\beta}_s \Phi_{\alpha}(\ner,\ner) =0\mbox{ and } \p^{\beta}_s \Phi_{n,\alpha}(\ner,\ner) =\lf\{\begin{array}{ccl} 1\mbox{ if }\beta=\alpha,\\
0\mbox{ if } \beta\neq\alpha,\end{array}\rg.
\end{equation}
 at~$\ner'=\ner$ for $|\beta|\leq 1$. As it turns out~\cite{perez2019planewave}, explicit analytical expressions for these functions can be derived. Indeed, letting $\bol\tau_j$, $j=1,2,$ denote linearly independent unit surface tangent vectors at $\ner\in\Gamma$ and further assuming that $\bol\tau_1\cdot\bol\tau_2= 0$ and $\bol\tau_1\times\bol\tau_2=\bnor$, we have that 
\begin{subequations}\label{eq:U2}\begin{align}
\Phi_{(0,0)}(\ner',\ner) &:=\frac{1}{k}\sin\lf(k\bnor\!\cdot\!(\ner'-\ner)\rg),\\
\Phi_{(1,0)}(\ner',\ner) &:=\frac{2}{k^2}\sin\lf(\frac{k\bnor}{\sqrt{2}}\!\cdot\!(\ner'-\ner)\rg)\sin\lf(\frac{k\bol\tau_1}{\sqrt{2}}\!\cdot\!(\ner'-\ner)\rg), \\
\Phi_{(0,1)}(\ner',\ner) &:=\frac{2}{k^2}\sin\lf(\frac{k\bnor}{\sqrt{2}}\!\cdot\!(\ner'-\ner)\rg)\sin\lf(\frac{k\bol\tau_2}{\sqrt{2}}\!\cdot\!(\ner'-\ner)\rg),
\end{align}\end{subequations}
satisfy~\eqref{eq:point_conditions} and, therefore,~\eqref{eq:interpolants} satisfies the point conditions~\eqref{eq:Iconds} for the interpolation orders we were looking for~\cite{perez2019planewave}.

\subsection{Numerical PWDI}\label{sec:alge}
An algebraic approach to find the coefficients $\{c_\ell(\ner)\}_{\ell=1}^{\ell=L}$ in the PWDI expansion~\eqref{eq:pwdi} at  a given point $\ner\in\Gamma$ is presented in this section.  Unlike the analytical approach, a collection of planewave directions $\{\bol d_\ell\}_{\ell=1}^{\ell=L}$ that are independent of the point $\ner\in\Gamma$ is used. While the desired interpolation orders~$M_j$, $j=1,2,$ and the number $L$ of planewave directions are parameters in the algorithm, the planewave directions themselves can be selected either randomly or uniformly from the unit sphere in three dimensions (an appropriate selection of planewave directions is provided in Sec.~\ref{sec:directions}).

In order to find the desired expansion coefficients one has to impose a number $D_1=(M_1+1)(M_1+2)/2$ of independent conditions~\eqref{eq:dir_cond} as well as $D_2=(M_2+1)(M_2+2)/2$ independent conditions~\eqref{eq:neu_cond},  which have to be satisfied exactly. Consequently, a solvable linear system for the coefficients could be produced provided the number of planewave directions satisfies  $L\geq D_1+D_2$. In order to form such a linear system, we proceed to sort the indices $\alpha=(\alpha_1,\alpha_2)$ satisfying $|\alpha|=\alpha_1+\alpha_2\leq \max\{M_1,M_2\}$ by introducing a bijective mapping $f:\{|\alpha|\leq \max\{M_1,M_2\}\}\to \{1,\ldots,\max\{D_1,D_2\}\}$.  Therefore, letting $\bold b(\ner)\in\C^{D}$, where $(\bold b(\ner) )_n = 0$, $1\leq n\leq D_1$ and $(\bold b(\ner) )_n = \p_s^{f^{-1}(n)}\varphi(\ner)$, $1\leq n\leq D_2$, we have that  conditions~\eqref{eq:Iconds} lead to the linear system 
\begin{equation}\label{eq:lin_sym1}
A(\ner)\bold c(\ner) = \bold b(\ner)
\end{equation}
for the coefficient vector~$\bold c(\ner)=[c_1(\ner),\ldots,c_{L}(\ner)]^T\in\C^{L}$,  where $A(\ner)$ is a $(D_1+D_2)\times L$ complex-valued matrix that depends on the planewave directions and the local geometry of the surface $\Gamma$ at the point $\ner$. Note that we have assumed in these derivations that the first $D_1$ rows of $A(\ner)$ correspond to the conditions~\eqref{eq:dir_cond} on $\Phi(\cdot,\ner)$, while the remaining $D_2$ rows correspond to the conditions~\eqref{eq:neu_cond} on $\Phi_n(\cdot,\ner)$ sorted according to the bijective mapping $f$. 

As it turns out the matrix $A(\ner)$ can be easily constructed at points where the surface is locally flat~\cite{perez2019planewave}. In fact, for interpolation orders $M_1=M_2=3$ the column of the $A(\ner)$ associated to the planewave direction $\bol d_\ell$ is given by $[\bold a_\ell(\ner), \tau_2\bold a_\ell(\ner)]^T$ where 
$\bold a_\ell=[1,\tau_1,\tau_2,\tau_1^2,\tau_1\tau_2,\tau_2^2,\tau_1^3,\tau_1^2\tau_2,\tau_1\tau_2^2,\tau_2^3]$ with $\tau_1=ik\bol d_\ell\cdot\bol\tau_1$, $\tau_2=ik\bol d_\ell\cdot\bol\tau_2$ and $\tau_3=ik\bol d_\ell\cdot\bnor$, and the bijective mapping defined by $f(0,0)=1$, $f(1,0)=2$ $f(0,1)=3$, $f(2,0)=4$, $f(1,1)=5$, $f(0,2)=6$, $f(3,0)=7$, $f(2,1)=8$, $f(1,2)=9$, and $f(0,3)=10$. 

\section{Implementation details}\label{sec:imp_details}
\subsection{Quadrature rule}\label{sec:quad_rule}
This section describes a straightforward quadrature rule for numerical integration over triangulated surfaces that will be used in  the approximation of the regularized surface integrals produced by the PWDI method.

First we focus on the double integral~\eqref{eq:single_layer_op}  which upon discretization of the surface $\Gamma$ into planar triangles becomes
\begin{equation}\label{eq:desc_triang}
\langle \psi,\mathsf S \varphi\rangle = \sum_{i=1}^{N_h}\int_{T_i}\psi(\ner)(\mathsf S\varphi)(\ner)\de s,\end{equation}
with 
\begin{equation}\label{eq:disc_sl}
(\mathsf S\varphi)(\ner) = \sum_{j=1}^{N_h}\int_{T_j}K(\ner,\ner')\de s'\quad 
\end{equation}
where the regularized integrand above  is given by \begin{equation*}
K(\ner,\ner') = G(\ner,\ner')\lf\{\varphi(\ner')-\Phi_n(\ner',\ner)\rg\}+ \frac{\partial G(\ner,\ner')}{\partial\bnor'}\Phi(\ner',\ner).
\end{equation*} 
 Notice that the term involving $\Phi(\ner,\ner)$ has been omitted here as it has been already assumed that  $\Phi(\cdot,\ner)$ interpolates the zero density at $\ner\in\Gamma$.

\begin{figure}[!t]
\centering
\includegraphics[scale=1]{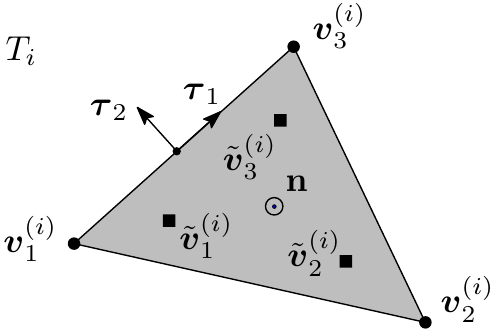}
\caption{Vertices (filled circles), unit normal (circle), and tangent vectors (arrows) associated to a mesh triangle~$T_i$.}
\label{fig:ext_triangle}
\end{figure}

 We recall now that  the construction of $\Phi$ requires both the scalar density $\varphi$ and the surface $\Gamma$ to be smooth in a neighborhood of $\ner\in \Gamma$. In order to fulfill these conditions we utilize the (interior) quadrature points~\cite{cowper1973gaussian}
 \begin{equation}
\tilde{\bol v}^{(i)}_{m}:=\sum_{\ell=1}^3 \frac{(1+3\delta_{m,\ell})}{6}\bol v^{(i)}_{\ell},\  1\leq m\leq 3,\ 1\leq i\leq N_h,\label{eq:quadrature_points}\end{equation}
to evaluate the integrals on the triangles $T_i$, where $\bol v^{(i)}_\ell$, $1\leq \ell\leq 3$,  denote the vertices of $T_i$ (see Fig.~\ref{fig:ext_triangle}). Application of this quadrature rule yields the following approximation
\begin{equation}\label{eq:quad_rule}
\langle \psi,\mathsf S \varphi\rangle \approx\sum_{i=1}^{N_h}\frac{A_i}{3}\sum_{m=1}^3\psi(\tilde{\bol v}^{(i)}_m)(\mathsf S\varphi)(\tilde{\bol v}_m^{(i)}),
\end{equation} of~\eqref{eq:desc_triang}, where $A_i$ is the area of  $T_i$. Given that the quadrature points $\tilde{\bol v}_m^{(i)}$ lie in the interior of $T_i$, we use for the computation of the planewave interpolant  $\Phi$ the unit normal $\bnor$ to $T_i$ and orthogonal unit vectors $\bol\tau_\ell$, $\ell=1,2,$ tangential to $T_i$, such as those shown in Fig.~\ref{fig:ext_triangle}.

The values of the single-layer operator at the quadrature points~$\tilde{\bol v}^{(i)}_m$, $1\leq m\leq 3$, which are needed in~\eqref{eq:quad_rule}, are approximated as
\begin{equation}\label{eq:desc_triang_on}
(\mathsf S \varphi)(\ner) =\sum_{i=1}^{N_h}\int_{T_i} K(\ner,\ner')\de s'\approx\sum_{i=1}^{N_h}\frac{A_i}{3}\sum_{m=1}^{3} K(\ner,\tilde{\bol v}^{(i)}_m)
\end{equation}
for $\ner\in\Gamma$, using the same interior point quadrature rule, but any other sufficiently accurate quadrature over triangles can be used. 

Similarly,  the regularized single-layer potential~\eqref{eq:reg_SL} is approximated as
\begin{equation}\label{eq:desc_triang_off}
(\mathcal S \varphi)(\ner) =\sum_{i=1}^{N_h}\int_{T_i} K_0(\ner,\ner')\de s'\approx\sum_{i=1}^{N_h}\frac{A_i}{3}\sum_{m=1}^{3} K_0(\ner,\tilde{\bol v}^{(i)}_m)
\end{equation}
for $\ner\in\R^3\setminus\overline\Omega$, where the regularized integrand in this case takes the form \begin{equation}\begin{split}
 K_0(\ner,\ner') =G(\ner,\ner')\lf\{\varphi(\ner')-\Phi_n(\ner',\ner_0)\rg\}+ \\ \frac{\partial G(\ner,\ner')}{\partial\bnor'}\Phi(\ner',\ner_0),\end{split}
\end{equation}
with the interpolation point $\ner_0\in \Gamma$  selected as explained in what follows. For any given target point $\ner$, we first find the surface triangle $T_{i^*}$ whose center is the closest to the point $\ner$. The interpolation point is then selected as $\ner_0=\tilde{\bol v}^{(i^*)}_{m^*}$ where $\tilde{\bol v}^{(i^*)}_{m^*}$ is the (interior) quadrature point in $ T_{i^*}$ that is the closest to~$\ner$. Accordingly, the construction of the planewave interpolant uses the unit normal and tangent vectors to $T_{i^*}$. Note that this choice of $\ner_0$ ensures that the limit~\eqref{eq:limit_Spot} holds for the discretized single-layer operator~\eqref{eq:desc_triang_on} and potential~\eqref{eq:desc_triang_off}.  The gradient of the regularized  single-layer potential---which is also needed for the computation of the off-surface operator $\mathcal E$---is approximated following the same procedure applied to $\nabla K_0$ instead of $K_0$.

Finally, we consider the case in which $\Gamma=\Gamma_1\cup\Gamma_2$ is a composite surface in the sense defined in Sec.~\ref{sec:sef_intersec}. The scalar density $\varphi$ in this case has two components: $\varphi_1$ and $\varphi_2$ defined on  $\Gamma_1$ and $\Gamma_2$, respectively.   Letting  $\mathcal S_j$ denote the regularized single-layer potentials defined by integration on the closed surface~$\Gamma_j$, $j=1,2$, we have that the whole potential naturally splits as $\mathcal S\varphi = \mathcal S_1\varphi_1+\mathcal S_2\varphi_2$ into two terms, each of which can be discretized and evaluated everywhere---in $\R^3\setminus\overline\Omega$ and $\Gamma$---following the procedure described above using separate meshes for $\Gamma_1$ and~$\Gamma_2$.  These potentials are also utilized to compute the double integrals for the construction of the Galerkin impedance matrix. In fact, letting $\psi_1$ and $\psi_2$ denote the components of $\psi$ defined on $\Gamma_1$ and $\Gamma_2$, we have 
\begin{equation}\begin{split}
\langle \psi,\mathsf S \varphi\rangle =&\langle \psi_1,(\mathcal S_2 \varphi_2)|_{\Gamma_1}\rangle+\langle \psi_2,(\mathcal S_1 \varphi_1)|_{\Gamma_2}\rangle+\\
&\langle \psi_1,(\mathcal S_1 \varphi_1)|_{\Gamma_1}\rangle+\langle \psi_2,(\mathcal S_2 \varphi_2)|_{\Gamma_2}\rangle,
\end{split}\end{equation}
where each of the integrals $\langle \psi_i,(\mathcal S_j\varphi_j)|_{\Gamma_i}\rangle$ over~$\Gamma_i$ ($i,j=1,2$) can be approximated using the interior point quadrature rule described above.

\subsection{Selection of planewave directions}\label{sec:directions}

As was discussed in Section~\ref{sec:alge} above, the high-order algebraic approach  for the construction of the planewave interpolant requires the explicit selection of $L\geq D_1+D_2$ planewave directions where, $D_1=(M_1+1)(M_1+2)/2$ and $D_2=(M_2+1)(M_2+2)/2$ depend on the interpolation orders $M_1\geq 0$ and $M_2\geq 0$. We have observed in numerical experiments that the minimal choice $L=D_1+D_2$ results in a square matrix $A(\ner)$ that is very ill conditioned for some points $\ner\in\Gamma$. Therefore, we recommend in general to select $L>D_1+D_2$ in order to sufficiently enrich the column space of~$A(\ner)$ so that its pseudoinverse, denoted by~$A^\dagger(\ner)$, becomes computable. In practice, large enough  $\mathcal O(M_1M_2)$ numbers of planewave directions selected from a ``uniform" spherical grid, give rise to numerically invertible matrices $A(\ner) A^*(\ner)$ from where the $A^\dagger(\ner)$ can be computed. In detail, the planewave directions for the construction of numerical PWDI interpolants used throughout  this paper are given by
$(\cos\theta_m\sin\phi_n,\sin\theta_m\sin\phi_n,\cos\phi_n)$ where $\theta_m=2\pi(m-1/2)/L_\theta$ for  $m=1,\ldots,L_\theta$ and $\phi_n=\pi (n-1/2)/L_\phi$ for $n=1,\ldots,L_\phi$, with $L=L_\theta\times L_\phi=2\times 2,4\times3,5\times4,6\times5$ for interpolation orders $\max\{M_1,M_2\}=0,1,2,3$, respectively.

\subsection{Tangential derivatives of RWG basis functions}\label{sec:RWG}
\begin{figure}[!t]
\centering
\includegraphics[scale=1]{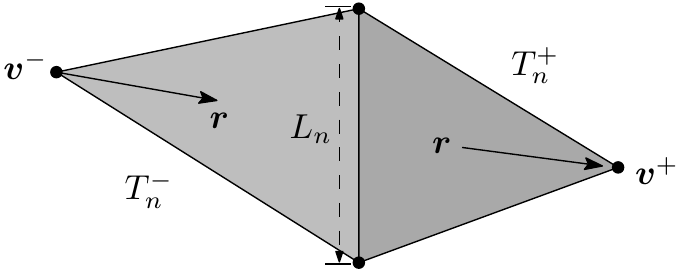}
\caption{Pair of triangles supporting the RWG basis function~$\bold f_n$ defined in~\eqref{eq:RWG_basis}.}
\label{fig:RWG}
\end{figure}

The RWG basis functions associated with the mesh edges are defined as~\cite{rao1982electromagnetic}
\begin{equation}
\bold f_n(\ner) := \lf\{\begin{array}{ccc} \displaystyle\pm\frac{L_n}{2A^\pm_n}(\bol v^\pm-\ner),&\ner\in T_n^\pm,\smallskip\\
\bol 0,&\ner\not\in T_n^\pm,
\end{array}\rg.
\label{eq:RWG_basis}\end{equation}
where $T_n^+$ and $T_n^-$ denote the triangles of areas $A_n^+$ and $A_n^-$, respectively, that  share the $n$-th edge of length $L_n$ (see Fig.~\ref{fig:RWG}).

The tangential derivatives of the RWG functions can be easily computed by differentiating~\eqref{eq:RWG_basis} and taking the dot product with the tangential unit vectors $\bol\tau_\ell^\pm$, $\ell=1,2$, associated to the corresponding triangles $T_n^\pm$. We thus have
\begin{equation}
(\p^\alpha_s \bold f_n)(\ner)=\lf\{\begin{array}{ccc} 
\displaystyle\mp\frac{L_n}{2A^\pm_n}\bol\tau^\pm_1(\ner),&\ner\in T_n^\pm,\ \alpha=(1,0),\smallskip\\
\displaystyle\mp\frac{L_n}{2A^\pm_n}\bol\tau^\pm_2(\ner),&\ner\in T_n^\pm,\ \alpha=(0,1),\smallskip\\
\bol 0,&\ner\not\in T_n^\pm\mbox{ or }|\alpha|>1,
\end{array}\rg.
\label{eq:diff_RWG}\end{equation}
and it also follows from~\eqref{eq:diff_RWG} that the tangential derivative of the surface divergence are given by 
$$
\p^\alpha_s(\nabla_s\cdot\bold f_n)(\ner) = \lf\{\begin{array}{rl} \displaystyle\mp \frac { L _ { n } } { A _ { n } ^ { \pm } },&\ner\in T_n^\pm,\ \alpha=(0,0)\smallskip\\
 0,&\ner\not\in T_n^\pm\mbox{ or }|\alpha|> 0.
\end{array}\rg.
$$

\section{Validation and examples} \label{sec:num_ex}
We present in this section a variety of numerical examples to validate and demonstrate the capabilities of the PWDI method. 

Letting $\elf^{\rm s}$ denote the MoM-computed scattered electric field  and letting $\elf^{\rm ref}$ denote the reference solution, the far-field and near-field errors are experimentally computed by means of the formula
\begin{equation}\label{eq:exp_error}
{\rm error} = \max_{\ner\in S}\lf|\elf^s(\ner)-\elf^{\rm ref}(\ner)\rg|/\max_{\ner\in S}\lf|\elf^{\rm ref}(\ner)\rg|
\end{equation}
with $S$ denoting  a set of sample target points. The set $S$ used in~\eqref{eq:exp_error} depends on the kind of error---in the far or near field---that is to be measured.  The far-field errors presented in the error plots in Sections~\ref{sec:single} and~\ref{sec:composite}, in particular, are  computed by taking $S$ as the set of mesh nodes corresponding to a large sphere of radius $r=100$m\footnote{All the physical quantities utilized in this section are in the MKS system.} centered at the origin, that encloses the surface~$\Gamma$ under consideration. 
\subsection{Single surfaces}\label{sec:single}
\begin{figure}[!t]
\centering
\includegraphics[scale=0.55]{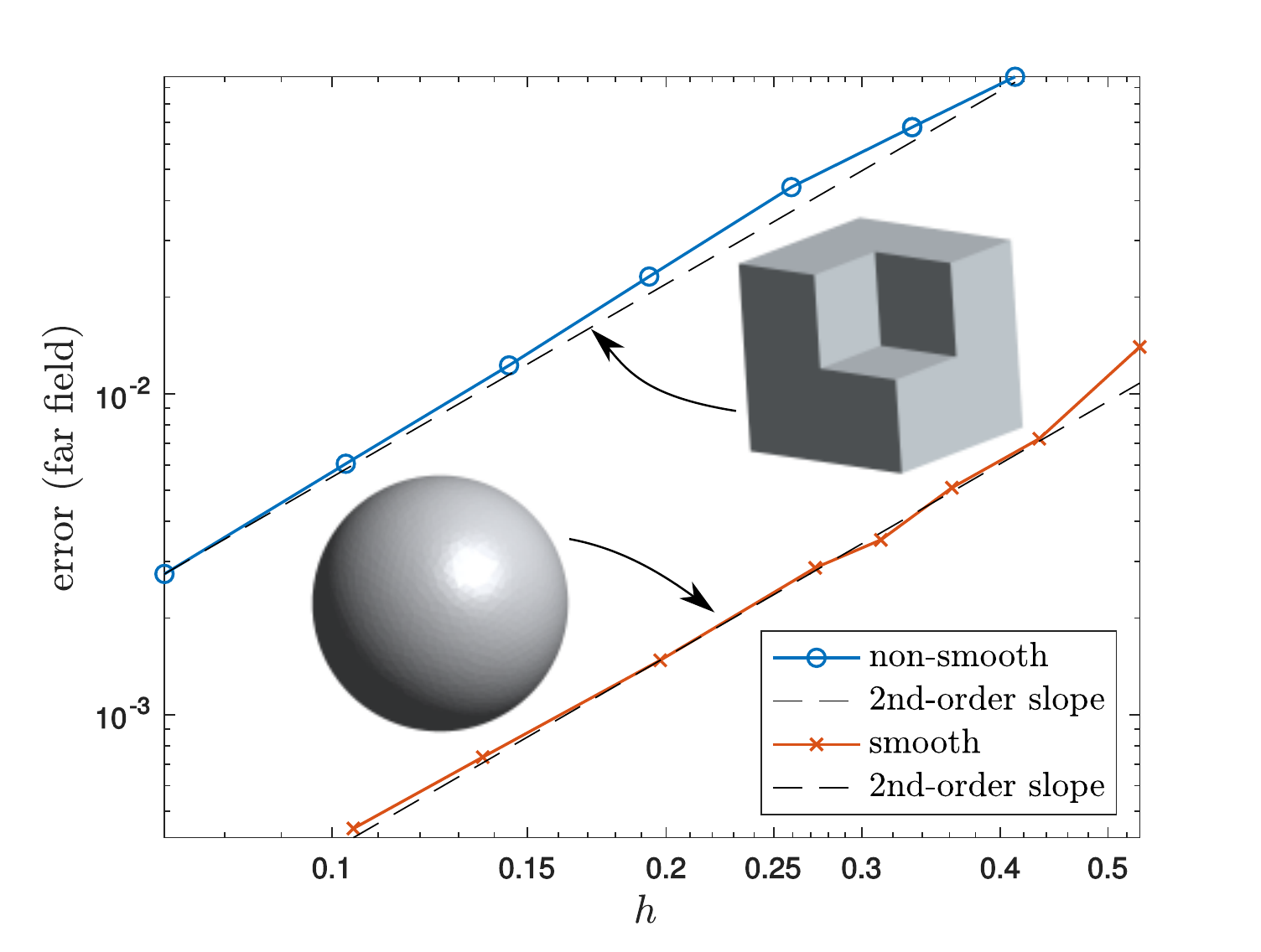}
\caption{Convergence of the far-field errors in the scattered field computed by means of the MoM using the analytical PWDI kernel-regularization procedure of Sec.~\ref{sec:analy_app} and the interior-point quadrature rule of Sec.~\ref{sec:quad_rule} .} 
\label{fig:FF_LO}
\end{figure}

The examples of this section have been designed to validate the proposed methodology when applied to the solution of standard problems of scattering by single PEC obstacles bounded by closed and simply connected surfaces. The frequency $f\approx75$MHz (corresponding to the wavenumber $k=0.5\pi\,{\rm rad/m}$ and  the wavelength $\lambda =4$m) is used in the examples of this section. We consider both a smooth surface and a more general Lipschitz surface featuring a reentrant corner at the origin. Specifically, these surfaces are respectively a sphere of radius~$1$m centered at the origin, and the boundary of the domain corresponding to a cube centered at the origin of side length~$1$m without the subdomain contained in the first (+++) octant (see inset  in Fig.~\ref{fig:FF_LO}).  The analytical PWDI approach of Section~\ref{sec:analy_app} is utilized here, but almost identical results are also obtained using the higher-order algebraic procedure presented in Section~\ref{sec:alge}. (This might be due because the Galerkin approximation errors dominate over the errors introduced by the numerical evaluation of the PWDI kernel-regularized integrals.) Figure~\ref{fig:FF_LO} displays the far-field errors obtained for various mesh sizes $h$ (in meters).  Clearly, the expected second-order convergence of the far-field errors is observed in both cases.   These errors were measured by means of~\eqref{eq:exp_error} with an exact closed-form solution used as the reference~$\elf^{\rm ref}$. Such an exact solution is manufactured by setting the dipole
$\elf^\inc(\ner) = -\nabla\times\{ G(\ner,\ner')\bol p\}$, located at a point $\ner'\in \Omega$ inside the closed surface, as the incident electric field. Indeed, it can be shown that the (unique) scattered electric field solution of the scattering problem is given by $\elf^s(\ner)= \nabla\times\{ G(\ner,\ner')\bol p\}$ ($\ner\in \R^3\setminus\Omega$) in this case. The  polarization vector $\bol p=(1,1,1)$ and the dipole location $\ner'= (-0.1,-0.1,-0.25)$ were used in both  smooth and non-smooth cases considered. 

\begin{figure}[!t]
\centering
\subfloat[Smooth surface]{\includegraphics[scale=0.55]{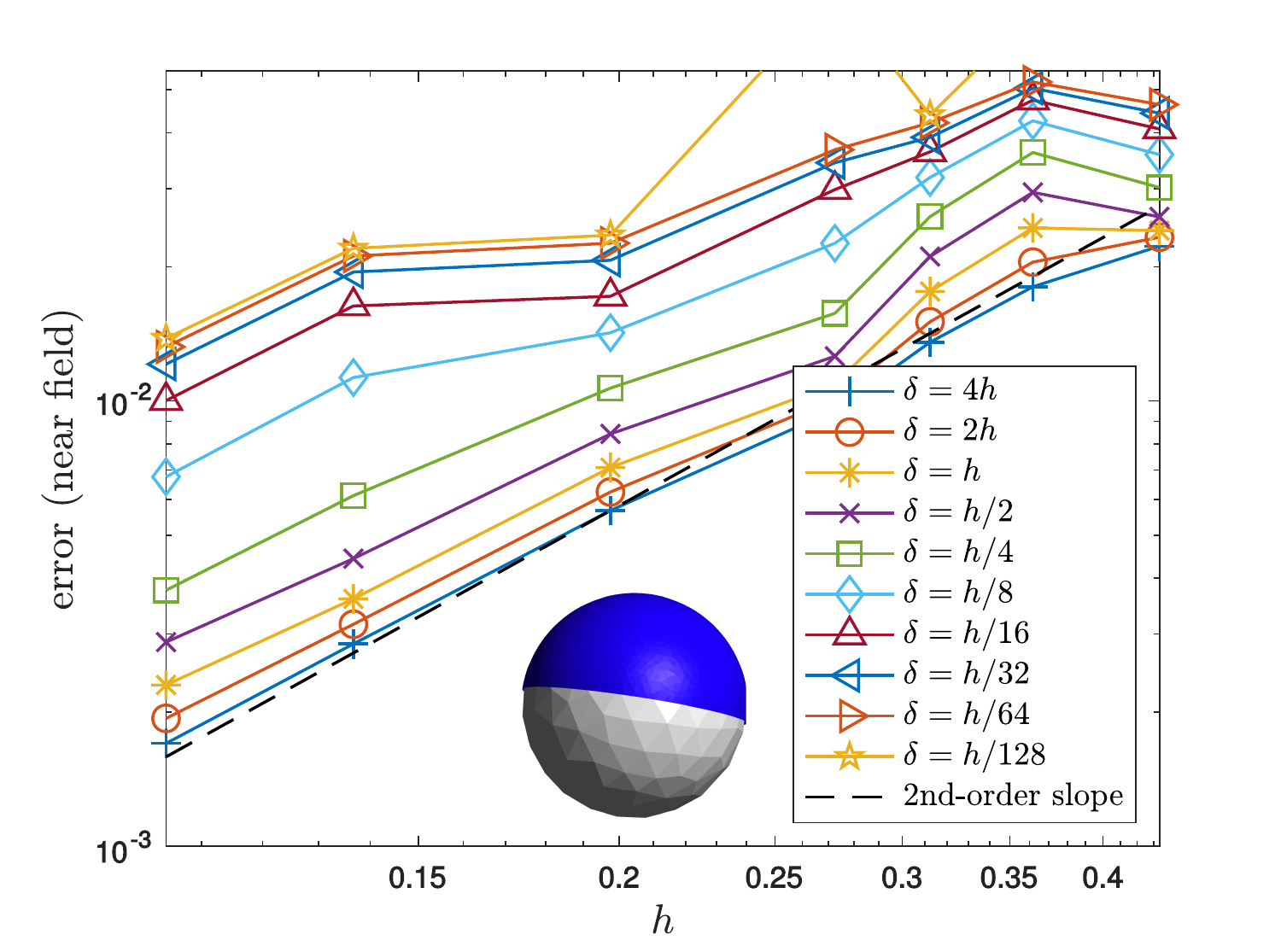}
\label{fig_first_case}}\vspace{-0.3cm}\\
\subfloat[Non-smooth surface]{\includegraphics[scale=0.55]{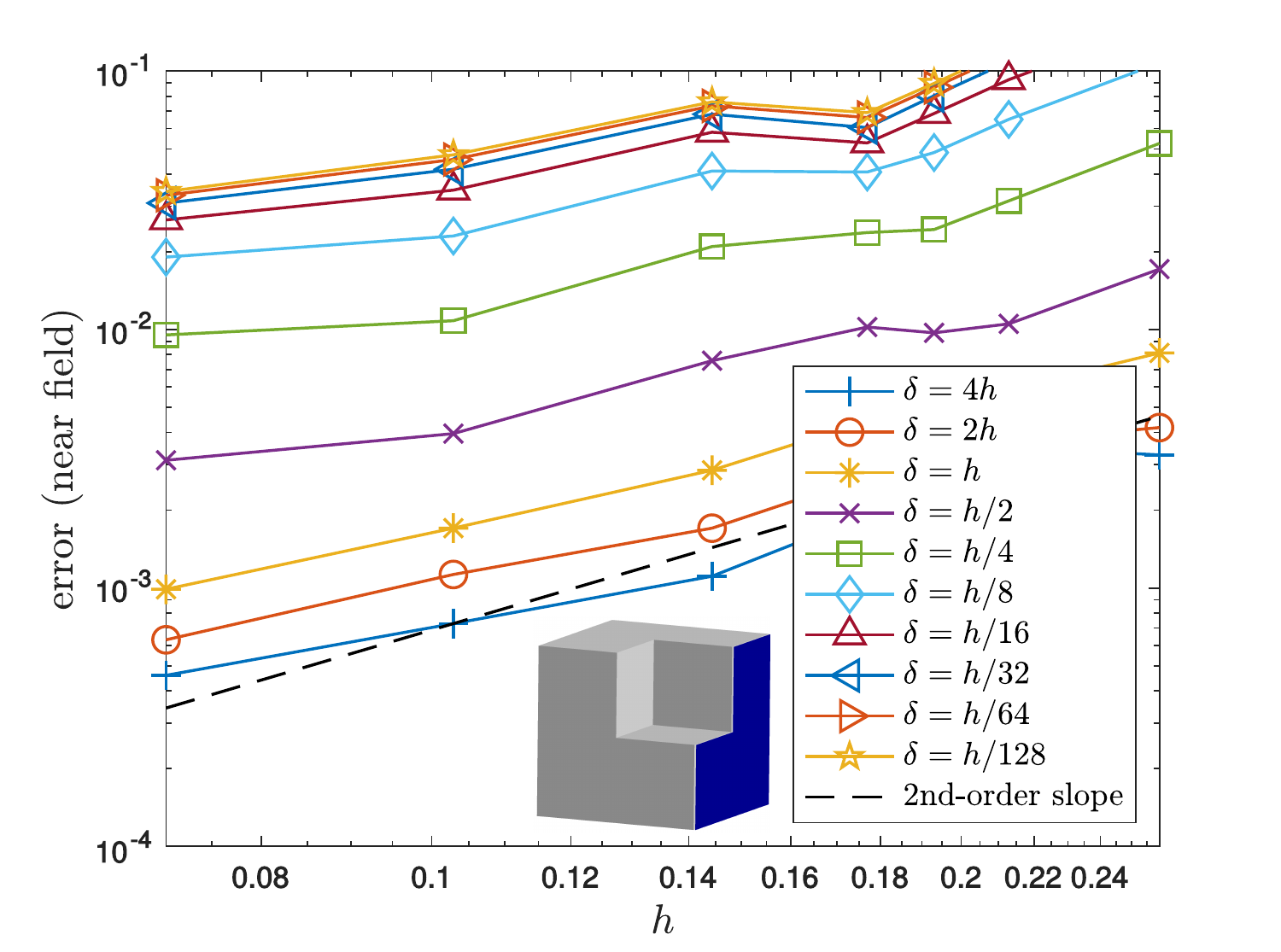}
\label{fig_second_case}}
\caption{Convergence of the near-fields errors in the solution of a scattering problem from a smooth and a non-smooth closed surface. The near-fields errors are empirically estimated using~\eqref{eq:exp_error} evaluating the fields at open surfaces $S$ that are parallel to the closed surfaces~$\Gamma$ under consideration. The evaluation surfaces~$S$, which are placed at a distance~$\delta$ from the closed surfaces, are depicted in blue in the inset figures.}
\label{fig:nf}
\end{figure}

Figure~\ref{fig:nf}, on the other hand,  presents the near-field errors corresponding to the same example problems, which were produced by means of~\eqref{eq:exp_error} with sample target points $S$ on evaluation surfaces that are ``parallel'' to $\Gamma$ and placed at a distance $\delta>0$ from $\Gamma$. These evaluation surfaces are depicted in blue in the inset of Fig.~\ref{fig:nf}. The higher-order interpolation procedure of Section~\ref{sec:alge}, with interpolation orders $M_1=M_2=3$, is used here. Notice that according to the estimates~\eqref{eq:control_ho_kernel_pot} and~\eqref{eq:control_ho_grad_kernel_pot} interpolation orders $M_1,M_2>1$ are needed in order to effectively regularize the off-surface EFIE operator $\mathcal E$ from where the scattered electric field is retrieved.  As is well-known, the accuracy of the MoM-produced off-surface EFIE operator depends not only on the distance  of the target point to the surface, but also on the local mesh size $h$ near the target point. Therefore, in order to account for the joint effect on the near-field error of these two variables,  we set the evaluation surfaces  at various distances $\delta$ from $\Gamma$ with $\delta$ being selected proportional to the mesh size $h$. As the results show, nearly second-order convergence of the near-field errors is achieved in all the cases considered. Significant accuracy deterioration is observed, however, for any fixed $h$ as $\delta$ becomes smaller. This deterioration is more substantial in the case of the non-smooth surface. This is explained by the fact that the limited smoothness of the EFIE solution $\bold J$ at and around corners and edges has a direct impact on the effectiveness of the interpolation procedure.

\subsection{Composite surfaces}\label{sec:composite}

\begin{figure}[!t]
\centering
\includegraphics[scale=0.55]{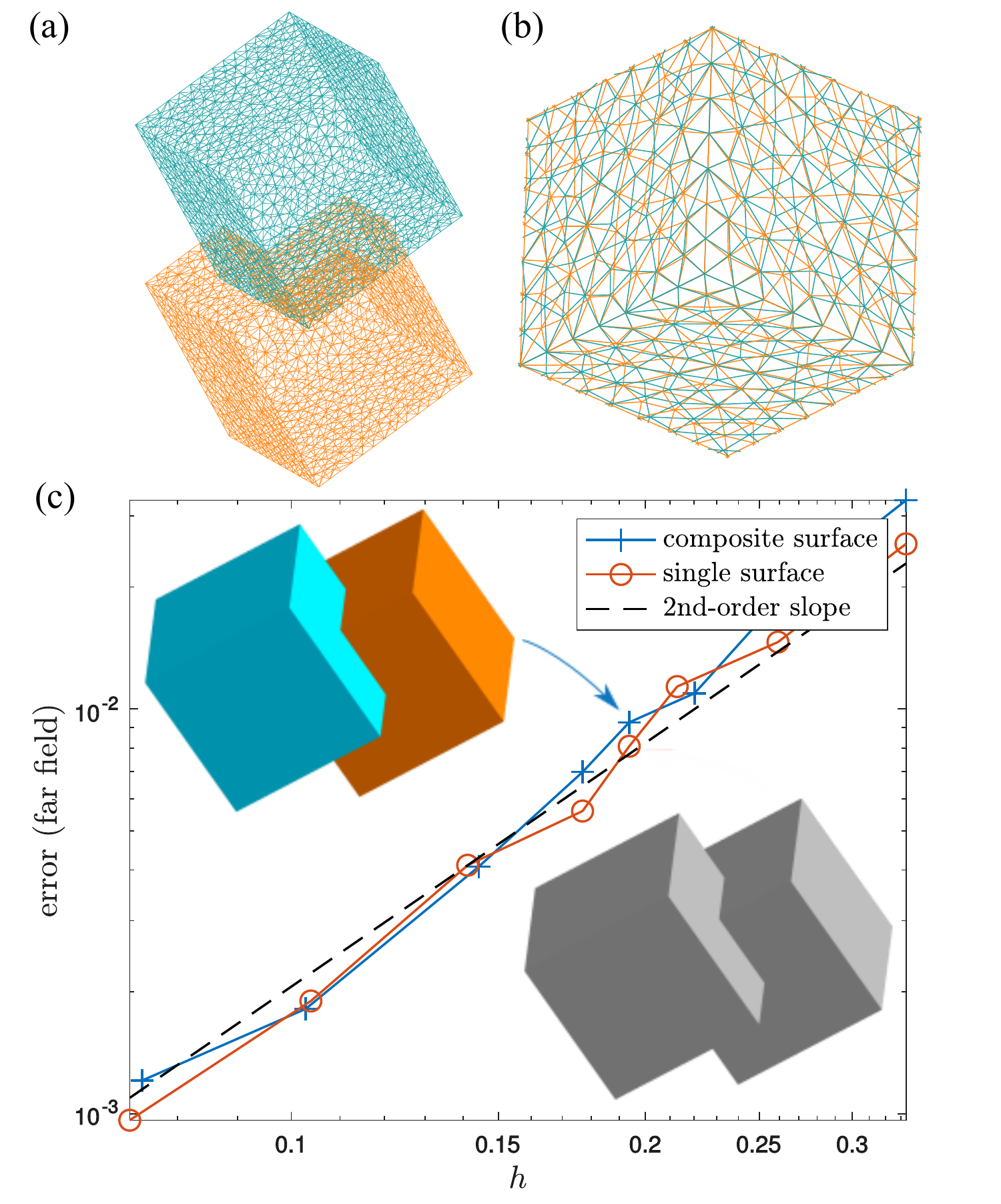}
\caption{Convergence of the far-fields errors obtained by means of the multiple-scattering EFIE formulation presented in Sec.~\ref{sec:sef_intersec}. (a)~An instance of the two non-conforming meshes used in the example. (b)~Zoomed view of the part where the two meshes overlap.  (c)~Far-field errors in the scattered fields produced by the proposed methodology applied to both the classical	 single-surface and the novel multiple-scattering EFIE formulations for various mesh sizes~$h$.}
\label{fig:convergence_cubes}
\end{figure}

In our next example, we consider a composite surface $\Gamma$ formed by two overlapping surfaces $\Gamma_1=\p\Omega_1$ and $\Gamma_2=\p\Omega_2$. Specifically, these are the non-smooth surface $\Gamma_1$ of the previous example---which is rendered in turquoise in the inset of Fig.~\ref{fig:convergence_cubes}c---and the boundary $\Gamma_2$ of a cube $\Omega_2$ of side length one contained in first octant---which is rendered in orange in the inset of Fig.~\ref{fig:convergence_cubes}c. We use here the  incident field~$\elf^\inc$ given by the superposition of two dipoles placed at $\ner_1'=(-0.1,0.1,-0.25)\in\Omega_1$ and $\ner_2'= (0.6,0.6,0.75)\in\Omega_2$ with polarizations $\bol p_2=(1,1,1)$ and $\bol p_2=(1,-1,1)$, respectively. As in the previous example, the associated  scattering problem has an exact closed-form solution given by $\elf^s(\ner) = \nabla\times\{ G(\ner,\ner_1')\bol p_1+  G(\ner,\ner_2')\bol p_2\}$, that was used in~\eqref{eq:exp_error} as the reference field to produce the error curves in Fig.~\ref{fig:nf}c. Both single- (standard) and multiple-scattering EFIE formulations---posed  on  $\p(\overline{\Omega_1}\cup\overline{\Omega_2})$ and on $\Gamma=\Gamma_1\cup\Omega_2$, respectively---are discretized and solved by means of the PWDI kernel-regularized MoM, which makes use of just one surface mesh of $\p(\overline\Omega_1\cup\overline\Omega_2)$ in the single-surface formulation, and  non-conformal overlapping surface meshes for $\Gamma_1$ and $\Gamma_2$ in the multiple-scattering formulation. Instances of the two non-conformal meshes, for $\Gamma_1$ and $\Gamma_2$, are shown in  Fig.~\ref{fig:convergence_cubes}a. The parts where the two meshes overlap are shown in more detail in Fig.~\ref{fig:convergence_cubes}b. The  convergence results  at $f\approx75$MGz ($k=0.5\pi\,{\rm rad/m}$) are presented in Fig.~\ref{fig:convergence_cubes}c which displays the far-field errors produced by the two approaches for various mesh sizes~$h$. The analytical PWDI procedure of Section~\ref{sec:analy_app} was used in this example. (The higher-order procedure of Section~\ref{sec:alge} produces almost identical results.) Nearly uniform meshes of all the surfaces involved where used in this example. (No mesh refinement was performed around edges or corners.) As can be observed in this figure, second-order convergence of the far-field errors is obtained as $h\to 0$ for both single- and multiple-scattering EFIE formulations and, furthermore, excellent agreement of the two is observed. 

\begin{figure}[!t]
\centering
\includegraphics[scale=0.55]{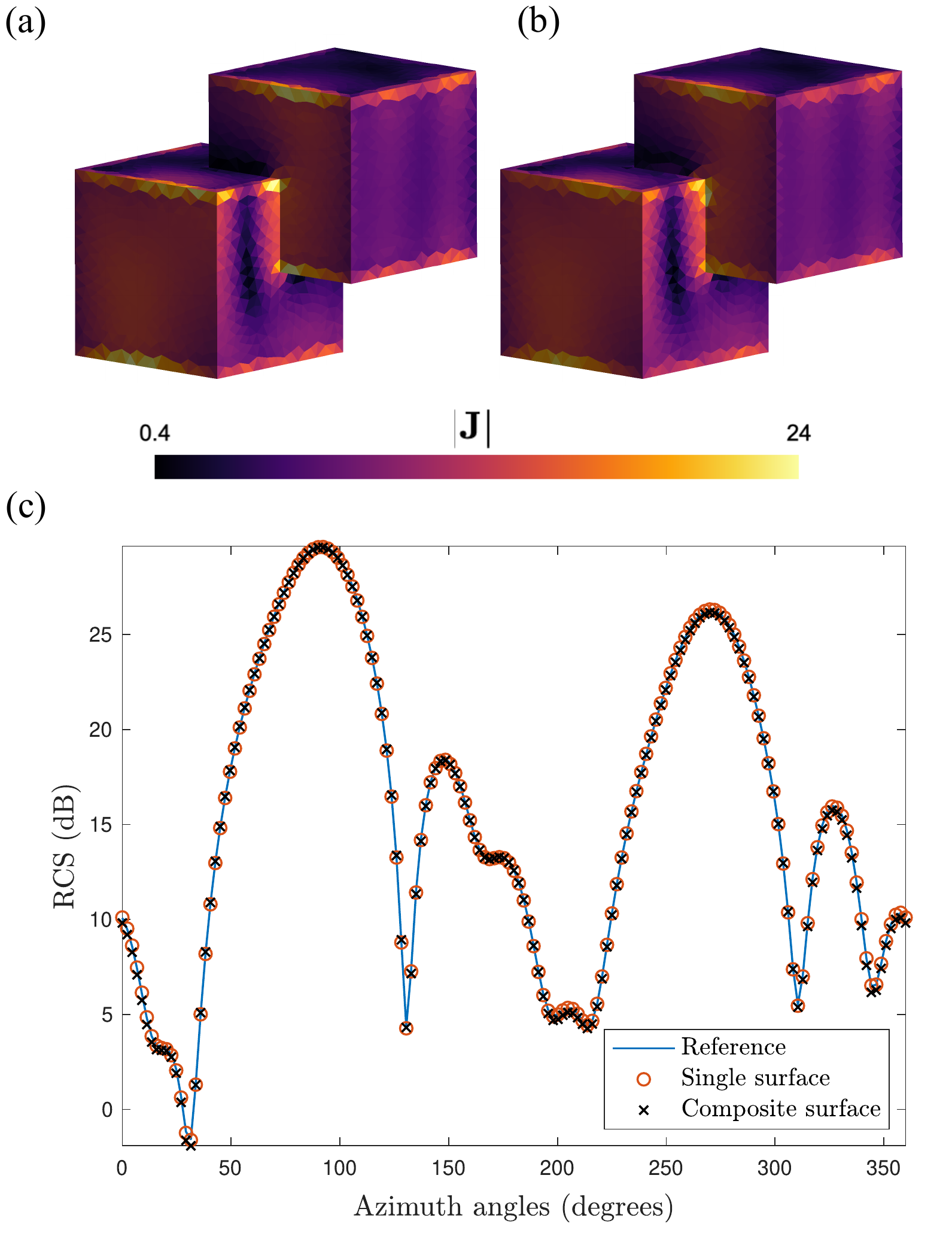}
\caption{Comparison of the single-surface and the multiple-scattering EFIE formulations for the solution of a problem of scattering. (a)~Magnitude of the surface currents corresponding to the single-surface formulation. (b)~Magnitude of the surface currents corresponding to the multiple-scattering formulation. (c)~RCSs at zero elevation angle. }
\label{fig:scatt_cubes}
\end{figure}

Keeping the  PEC structure of the previous example, we  consider now the problem of scattering of the planewave $\elf^\inc(\ner) = (\bol p\times\bol d)\e^{ik\bol d\cdot \ner}$ with $\bol p =(0,0,1)$, $\bol d=(0,1,0)$ and  $f\approx 0.3$GHz ($k=2\pi\,{\rm rad/m}$, $\lambda=1$m). Once again, for validation purposes,  the two single- and multiple-scattering EFIE formulations are used, wherein nearly uniform meshes of size $h\approx 0.1{\rm m}= \lambda/10$ are utilized. The resulting surface currents are shown in magnitude format in~Figs.~\ref{fig:scatt_cubes}a and~\ref{fig:scatt_cubes}b corresponding to the EFIE solutions obtained using single conforming and composite non-conforming  surface meshes, respectively. Note that the large current densities around the edges of the structure are well  captured by the two solution approaches. The associated radar cross sections (RCSs) are shown in Fig.~\ref{fig:scatt_cubes}c at zero elevation angle, where they are also compared against a reference RCS obtained using a significantly refined single conforming surface mesh ($h=0.075$m).  As can be seen in that figure, the three RCSs are almost indistinguishable.  

In order to further validate the proposed PWDI kernel-regularization procedure, we consider an even more challenging PEC structure involving three touching spheres; $\Gamma_1$, $\Gamma_2$ and $\Gamma_3$ of radii $r_1=0.5{\rm m}$, $r_2=0.4{\rm m}$ and $r_3=0.6{\rm m}$ that are  centered at $\bol c_1=(-0.5,0,0)$, $\bol c_2=(0.4,0,0)$  and $\bol c_3\approx (0.0667,0.9428,0.0)$, respectively.  We first estimate the numerical errors using~\eqref{eq:exp_error} by manufacturing an exact solution of the scattering problem. This is done in this case by placing dipoles inside each one of the corresponding spheres, at $\ner'_1=(-0.45,0.05, 0.125)$, $\ner'_2= 0.36,-0.04,-0.1)$ and $\ner'_3=(0.1267,1.0028,0.06)$, with associated polarizations $\bol p_1=(1,-1,1)$, $\bol p_2=(1,1,1)$ and $\bol p_3=-\bol p_2$.  As expected, the empirically estimated far-field errors, which are presented in Table~\ref{tab:1}, exhibit 
second-order convergence as $h\to 0$. The three meshes used in  each one of the examples reported in the table are approximately of the same size $h$. No mesh refinement of any kind was used. 

\begin{figure}[!t]
\centering
\includegraphics[scale=0.54]{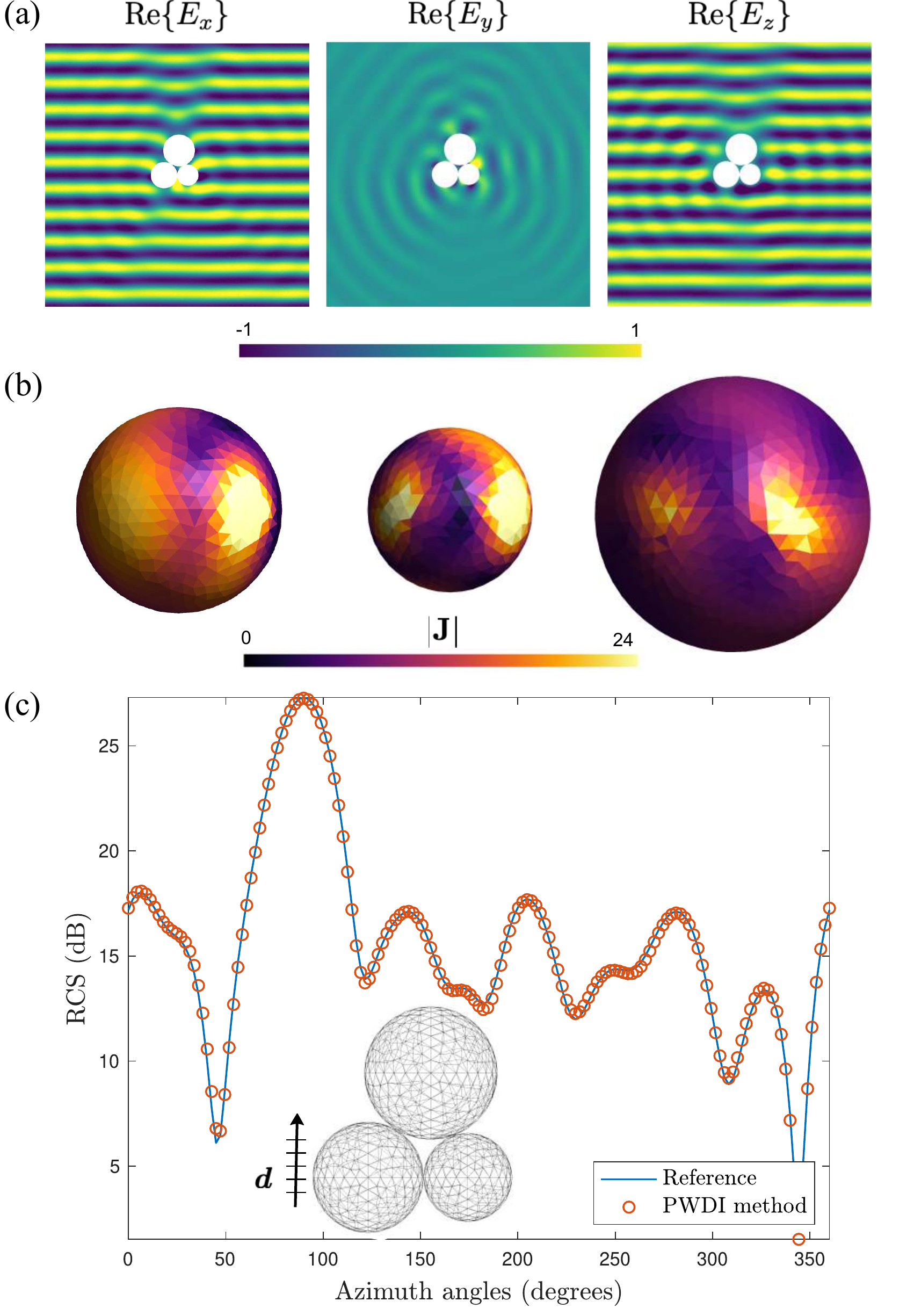}
\caption{Electric field and surface currents corresponding to the scattering of a planewave from a PEC structure comprising three touching spheres, computed using the PWDI method in conjunction with the MoM. (a)~Real part of the three components of the total electric field in the $xy$-plane. (b)~Magnitude of the surface currents on each of the spheres comprising the composite surface. (c)~RCSs at zero elevation angle.} 
\label{fig:scatt_sphere}
\end{figure}

Once we have validated the effectiveness of the PWDI procedure for this challenging PEC structure, we  move on to consider a more realistic scattering problem in which a planewave $\elf^\inc(\ner) = (\bol p\times\bol d)\e^{ik\bol d\cdot \ner}$, with $\bol p =(1,1,1)$, $\bol d=(0,1,0)$ and $f\approx 0.3$GHz ($k=2\pi\,{\rm rad/m}$, $\lambda = 1$m), illuminates the PEC structure. The results for this case are presented in Fig.~\ref{fig:scatt_sphere}. Surface meshes of approximately the same size $h\approx0.1{\rm m}=\lambda/10$ are used in this example. Figure~\ref{fig:scatt_sphere}a  shows the real part of the three Cartesian components of the total electric field $\elf=\elf^\inc+\elf^s$ on the plane containing the centers of the three spheres, which were produced by direct evaluation of the off-surface operator $\mathcal E$ after being regularized by the algebraic PWDI procedure of Sec.~\ref{sec:alge} with $M_1=M_2=3$. As expected, a weak shadow appears in the wake of the PEC structure. Figure~\ref{fig:scatt_sphere}b, on the other hand, displays  the surface currents (in magnitude format) at and around the touching points on each one of the three spheres.  Finally, Fig.~\ref{fig:scatt_sphere}c shows the associated RCSs together with a reference RCS, which was produced using finer meshes of size $h=0.075$m of the three spheres $\Gamma_1$, $\Gamma_2$, and $\Gamma_3$.

\begin{table}[h]

 \begin{center}

   \caption{Far-field errores resulting from a convergence test for the PEC structure shown in Fig.~\ref{fig:scatt_sphere} composed of three touching spheres.
}\label{tab:1}

   \begin{tabular}{c| c c c c cccc}
     \hline
   $h$ $(\times 10)$ & $0.69$ &$1.01$  & $1.37$ & $1.64$ & $1.87$ & $2.17$ & $2.67$ & 4.26\\
        \hline
  error (\%)  & 0.70 & 1.36   & 2.35  & 2.94   & 3.99 &5.78 &6.99 &24.8 \\
     \hline 
   \end{tabular}
 \end{center}
\end{table}

\begin{figure}[!t]
\centering
\includegraphics[scale=1.22]{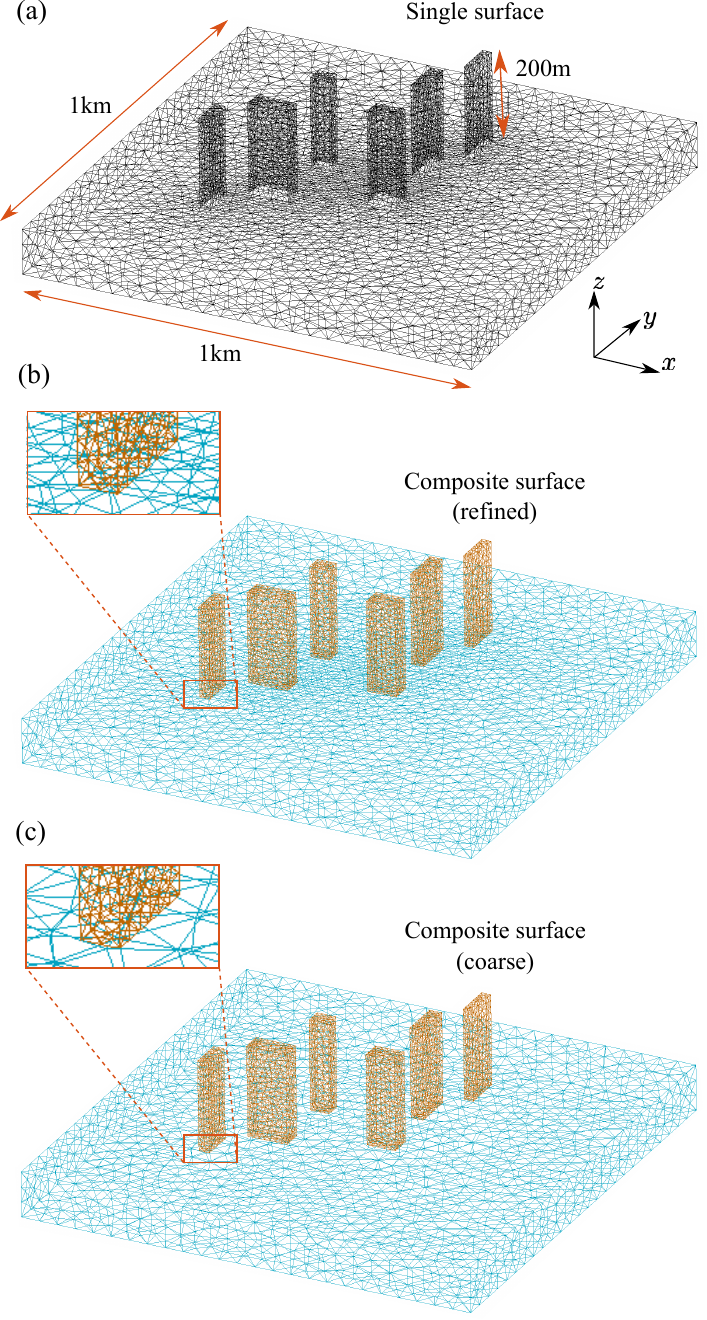}
\caption{Triangular surface meshes of a city-like multiscale structure. (a)~Mesh of a single surface representing the whole structure. (b)-(c)~Two non-conformal meshes representing the two parts of the structure; the small-scale building-like blocks (depicted in orange) and the large-scale ground-like structure (depicted in light blue). The ground mesh displayed in (b) is locally refined around the base of the building so as to match the mesh size of the small-scale structures.}
\label{fig:city_meshes}
\end{figure}

\begin{figure}[!t]
\centering
\subfloat[]{\includegraphics[scale=0.55]{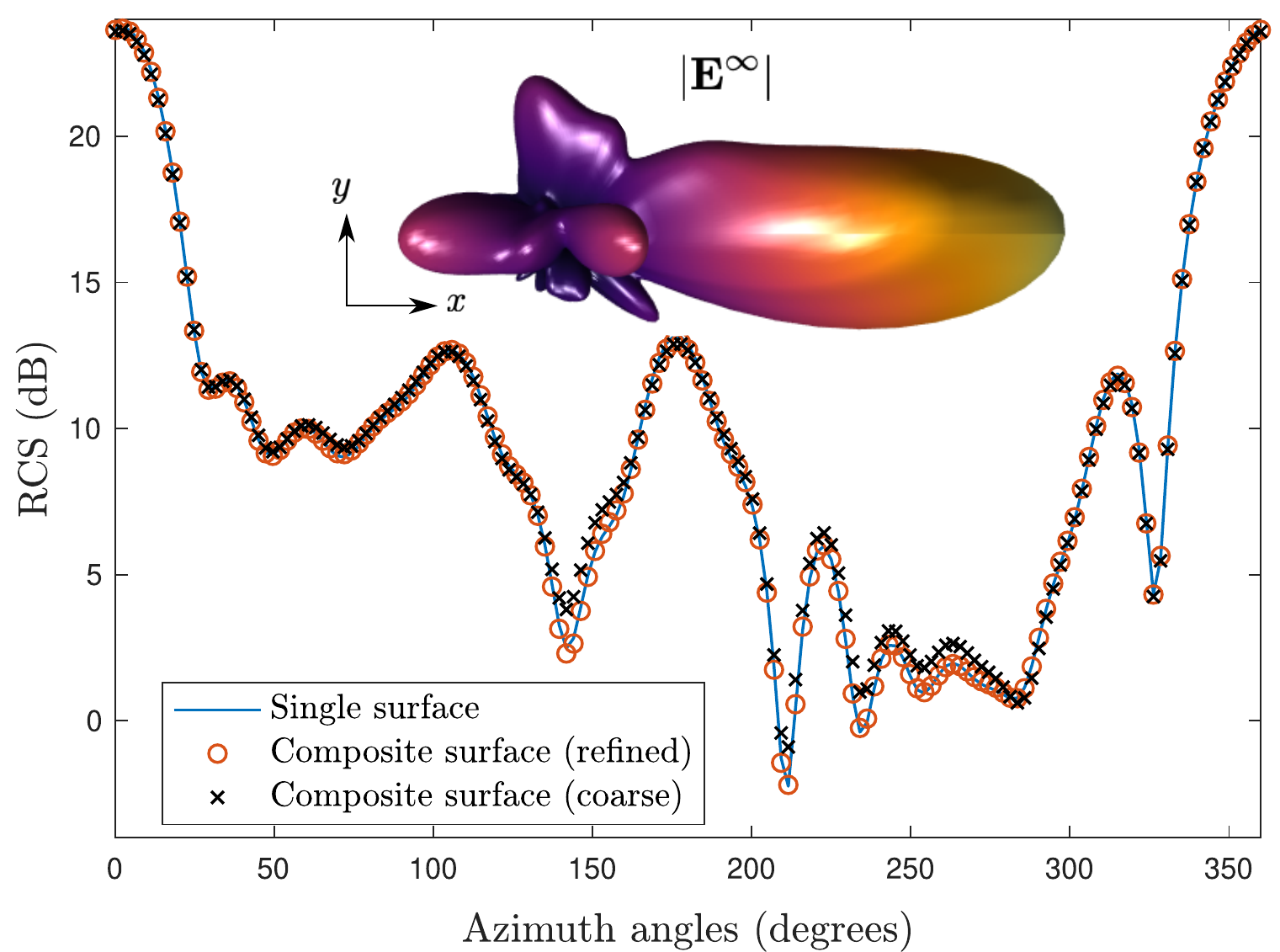}\label{fig:city_az}}\vspace{-0.2cm}\\
\subfloat[]{\includegraphics[scale=0.55]{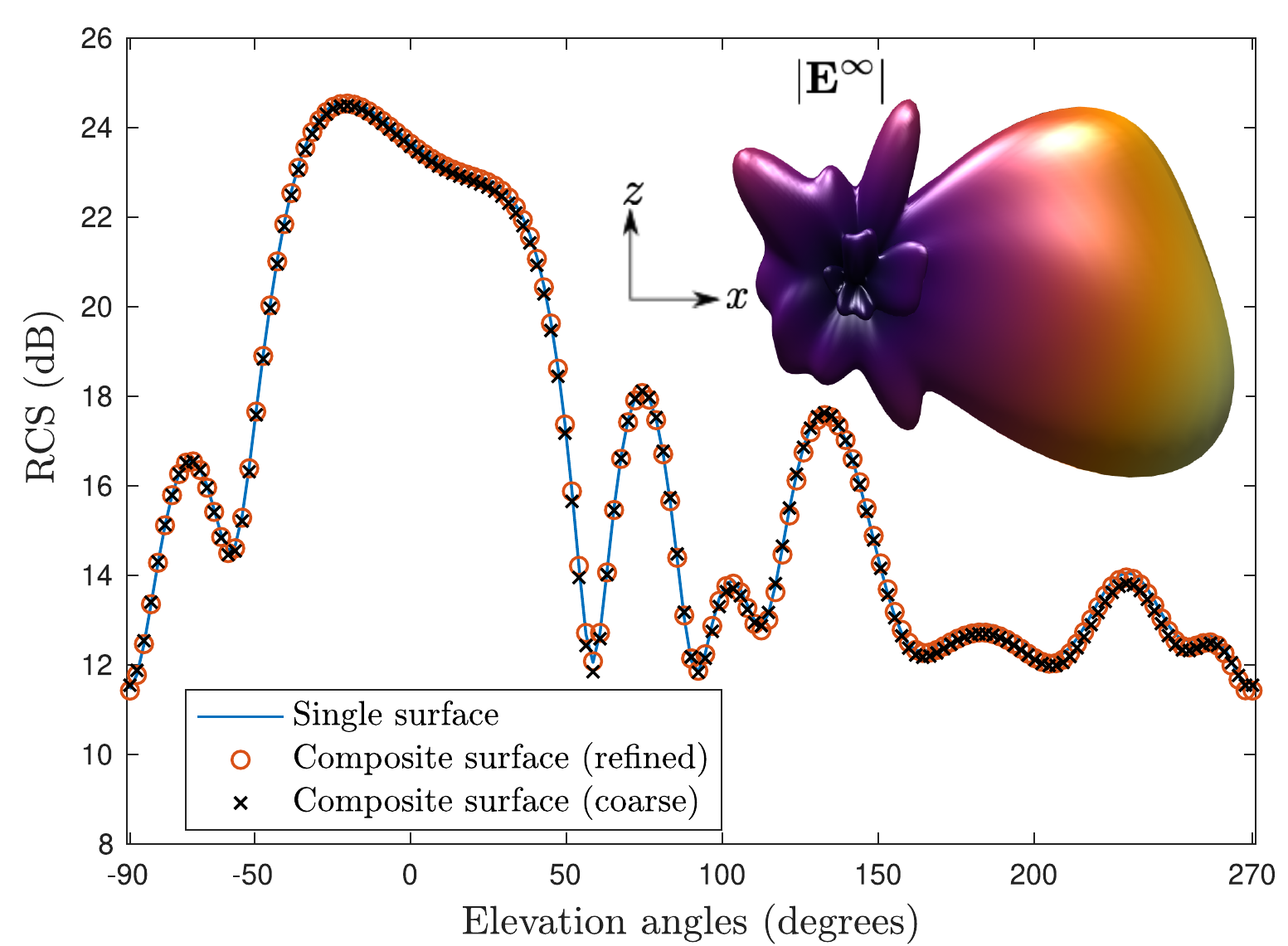}\label{fig:city_el}}
\caption{RCSs corresponding the solution of the problem of scattering of a planewave from the city-like PEC structure displayed in Fig.~\ref{fig:city_meshes}, at zero elevation angle (a), and at  zero azimuth angle (b). The inset figures display corresponding views of the three-dimensional far-field pattern.} 
\label{fig:RCS_city}
\end{figure}

\begin{figure}[!t]
\centering
\subfloat[]{\includegraphics[scale=0.55]{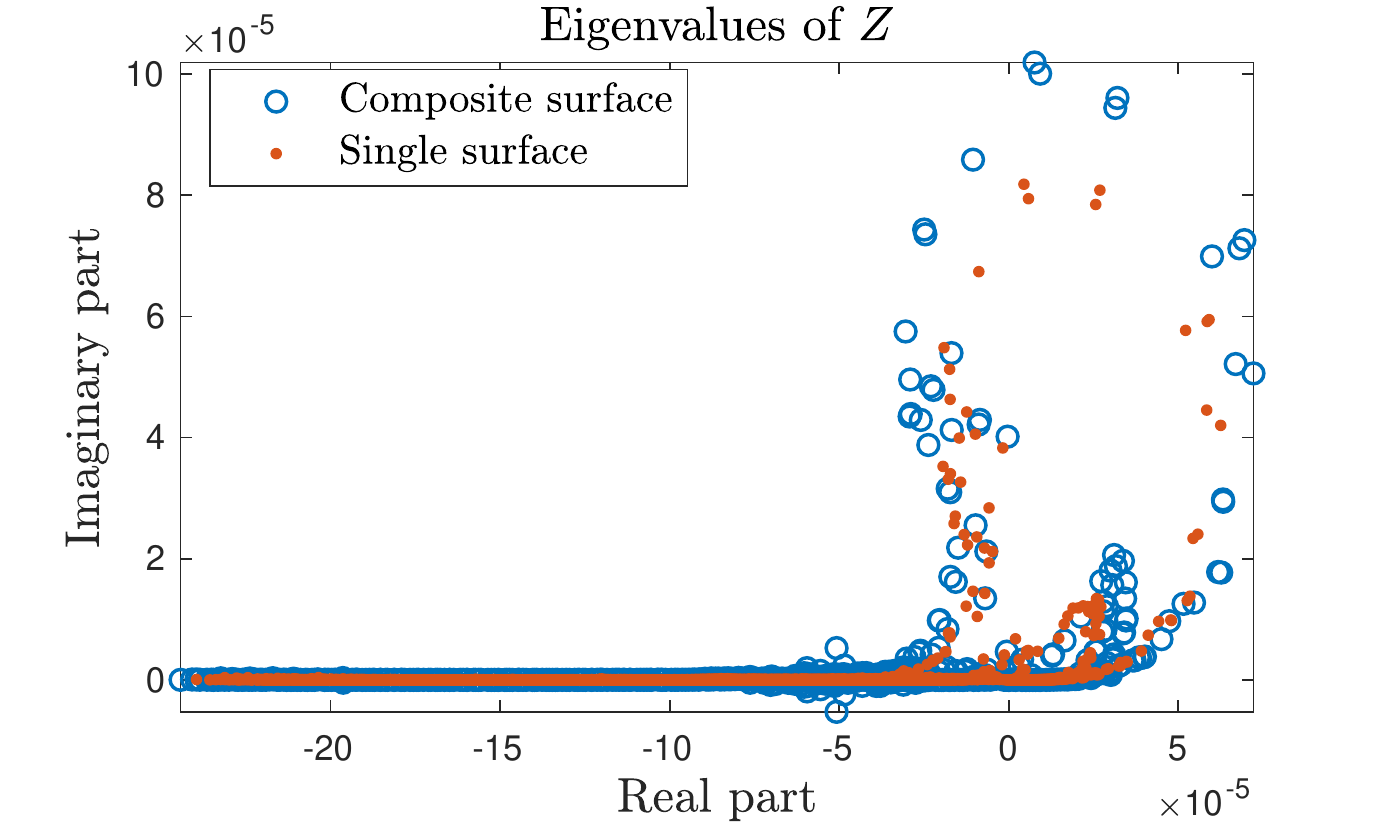}\label{fig:eigsSingle}}\\
\subfloat[]{\includegraphics[scale=0.55]{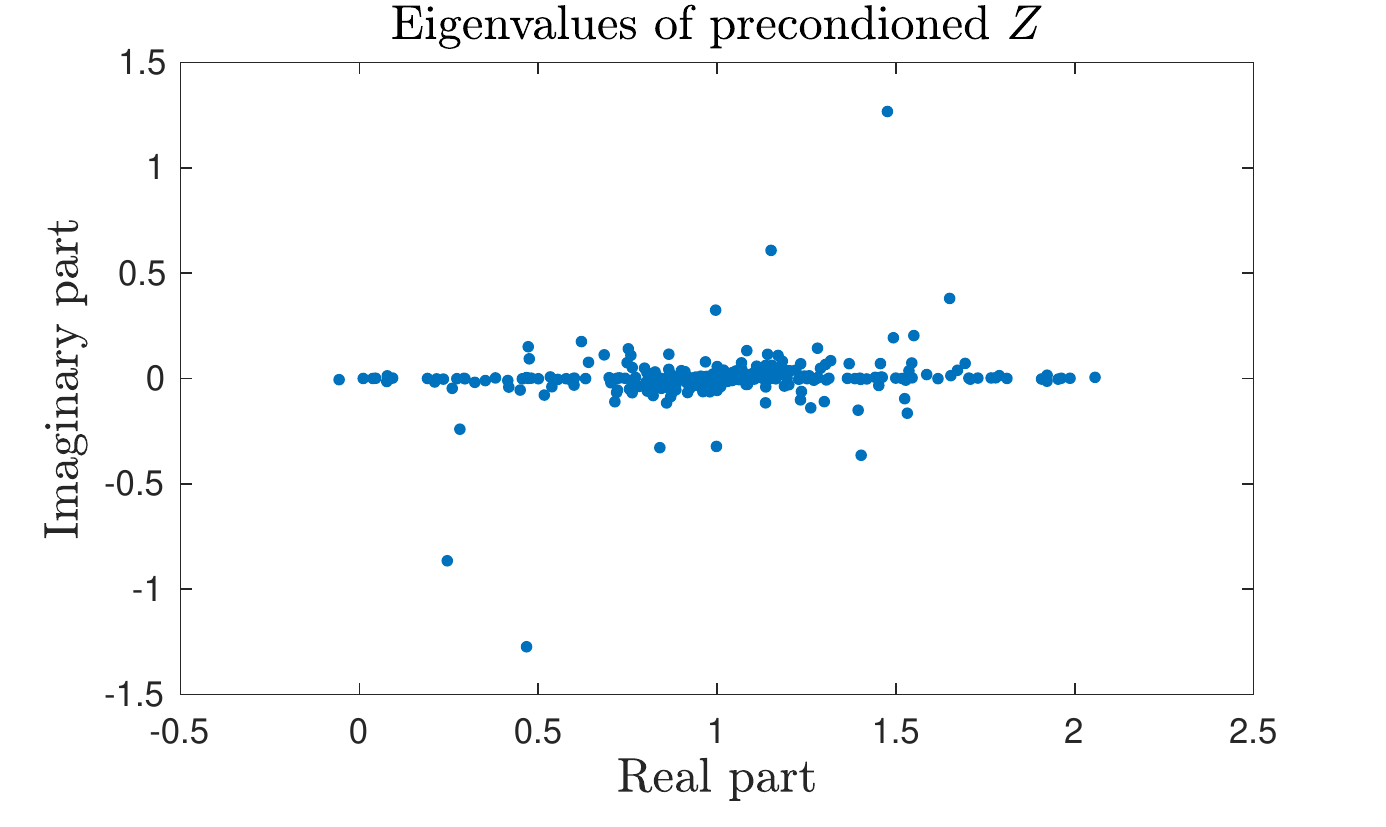}\label{fig:eigsComp}}
\caption{(a) Eigenvalues of the impedance matrices $Z$ corresponding to the composite- and single-surface EFIE formulations for the geometry displayed in Fig.~\ref{fig:city_meshes}. (b)~Eigenvalues of preconditioned impedance matrix corresponding the composite surface.}
\label{fig:eigs}
\end{figure}

\begin{figure}[!t]
\centering
\subfloat[Innovation Center building at PUC, Chile. Credit: ELEMENTAL (Nina Vidic).]{\includegraphics[scale=0.3]{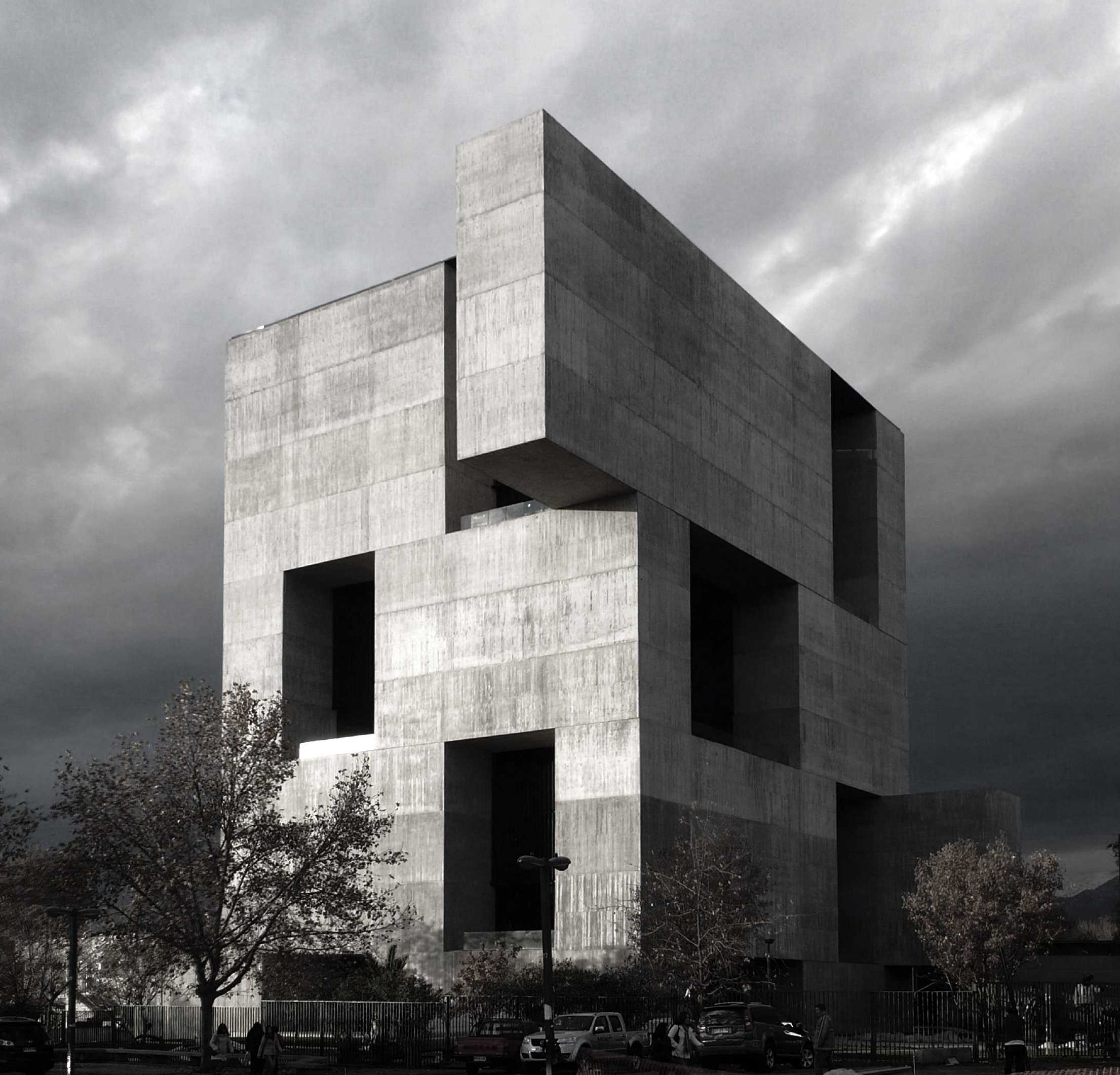}\label{fig:IC_photo}}\medskip\\
\subfloat[Composite-surface model of the Innovation Center building using non-conforming meshes. Each one of the sixteen rectangular blocks making up its intricate facade was meshed separately using  Gmsh software~\cite{geuzaine2009gmsh}. ]{\includegraphics[scale=0.33]{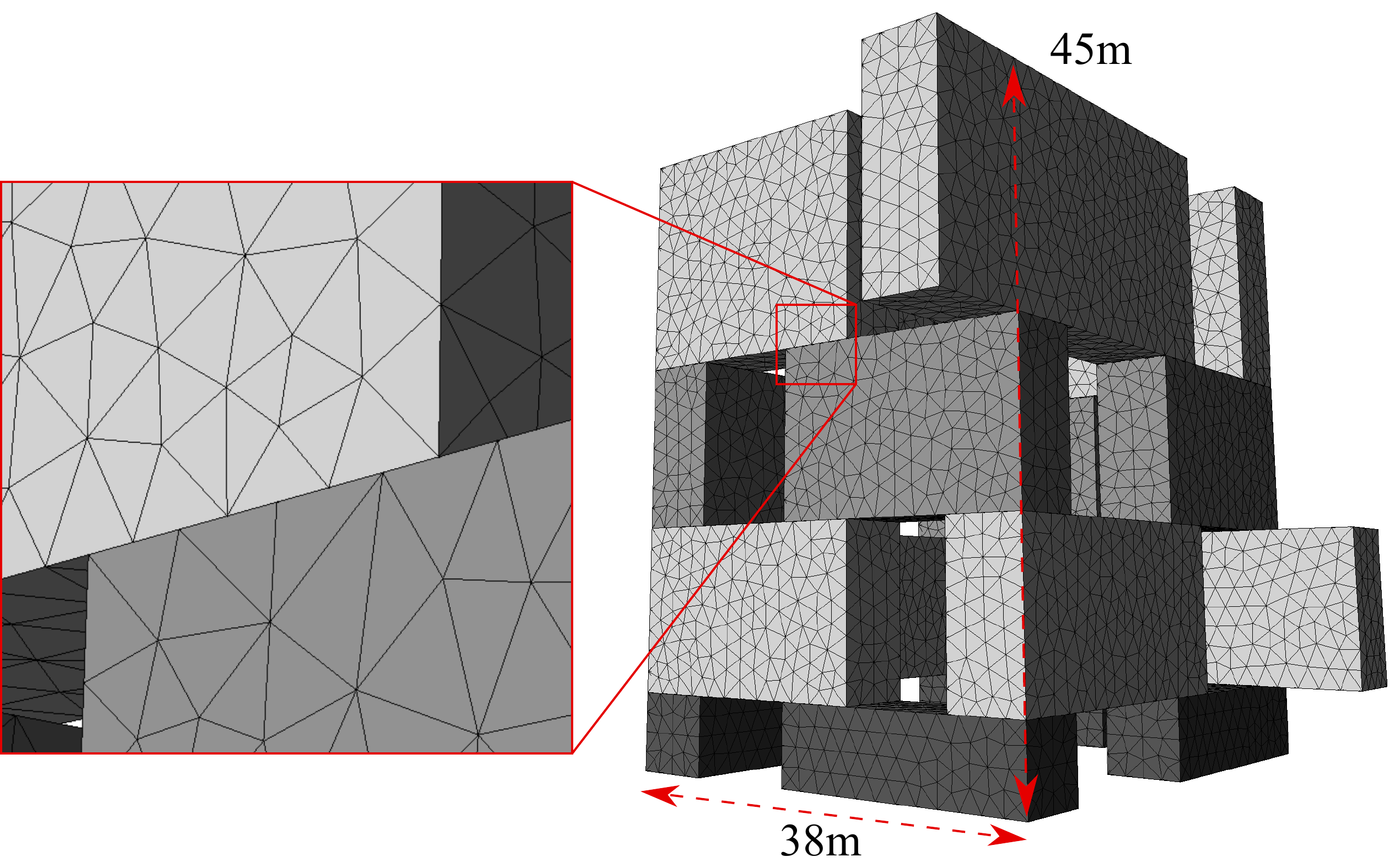}\label{fig:IC_meshes}}\\
\subfloat[Real part of total electric field on the middle cross section ($z=22.5$m) of the building. ]{\includegraphics[scale=0.2]{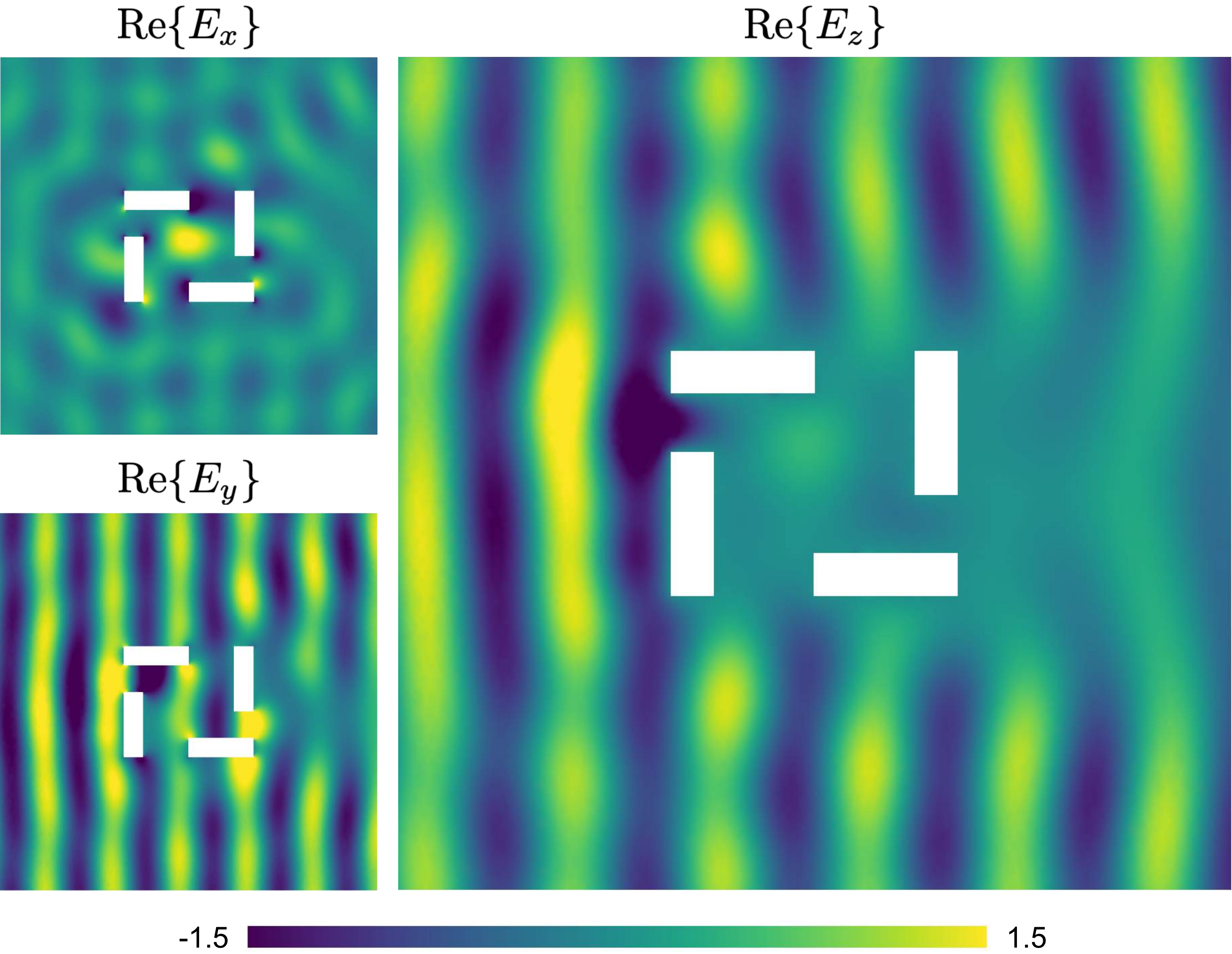}\label{fig:IC_NF}}
\caption{Example of an actual structure whose surface can be easily modeled using a composite non-conforming surface mesh consisting of sixteen rectangular blocks.}
\label{fig:IC}
\end{figure}

\subsection{Other examples}
This section is devoted to demonstrate the capabilities of the proposed PWDI methodology at dealing with scattering by more realistic structures, for which the multiple-scattering EFIE formulation can be remarkably advantageous. It includes two example problems corresponding to a multiscale city-like PEC structure, and another PEC structure whose complex shape is identical to that of the Anacleto Angelini Innovation Center Building at PUC's San Joaqu\'in campus in Santiago, designed by the Pritzker prize winner architect, Alejandro Aravena. 

\subsubsection{A multiscale object} We start off by considering a city-like PEC structure featuring six small-scale building-like structures, which are shown in  Fig.~\ref{fig:city_meshes}. The composite surface EFIE formulation enables in the case the solution of the problem using separate non-conformal meshes for the ground structure and for each one of the buildings, with mesh sizes adjusted according to the individual subdomain scales. This approach has in principle three main advantages over the standard single-surface EFIE formulation. Firstly,  it significantly reduces the overall number of  degrees of freedom (and, consequently, reduce the size of the system matrix), if no local refinement of the large-scale surface is used around the edges at the base of the smaller-scale subdomains (buildings). Secondly,  it significantly simplifies the mesh processing of the geometry, as  small-scale subdomains can be meshed separately and then be placed wherever is needed on the large-scale surface. And thirdly, the technique can be attractive for repetitive modeling and simulation, as it allows to effectively separate geometrical details of interest from the rest of the geometry of the problem. Consequently, the diagonal blocks of the system matrix, corresponding to self-interactions of the composite surface parts, do not need to be recomputed when the individual surface parts undertake a rigid (Euclidian) transformation.  

We solve here the  problem of scattering of a plavewave $\elf^\inc (\ner)=(\bol p\times \bol d)\e^{ik\bol d \cdot \ner} $ with $f\approx 0.6$MHz ($k=2\pi/500\,{\rm rad/m}$, $\lambda=500\,$m), $\bol d=(4,0,-1)/\sqrt{17}$ and $\bol p=(1,1,1)$, using both the standard and the multiple-scattering EFIE formulations. The algebraic PWDI kernel-regularization procedure with orders $M_1=M_2=3$ was used in the numerical discretization of the two formulations; almost identical results are obtained using the analytical PWDI procedure. Two different meshes for the ground structure are utilized; a refined mesh around the base of the buildings, shown in Fig.~\ref{fig:city_meshes}b, and a coarse mesh, irrespective of the presence of the buildings, that is shown in Fig.~\ref{fig:city_meshes}c. The standard EFIE formulation, on the other hand,  is discretized using the (conforming) mesh shown in Fig.~\ref{fig:city_meshes}a. 

The  RCSs resulting from the solution of each one of the problems comprised in this example are presented in Fig.~\ref{fig:RCS_city}.  The RCSs for the azimuth angles at zero elevation angle are shown in Fig.~\ref{fig:city_az}, while the RCSs for the elevation angles at zero azimuth angle are shown in Fig.~\ref{fig:city_el}. Corresponding views of the three-dimensional   far-field pattern $\elf^\infty(\hat\ner) = 4\pi\lim_{|\ner|\to\infty}|\ner|^2|\elf^s(\ner)$, with $\hat\ner=\ner/|\ner|$, are displayed in the inset of Figs.~\ref{fig:city_az} and~\ref{fig:city_el}. As can be observed in these figures,  a good agreement among the three RCSs is achieved, though slightly more accurate results are obtained using the locally refined mesh of the ground structure, especially around the $250^{\circ}$ and $0^{\circ}$ azimuth  and elevation angles, respectively. Given the scale differences, the resulting surface currents around the base of buildings might not be well resolved when the  coarse ground mesh is used, and this ends up introducing errors that are mitigated using local refinement. It is worth mentioning that although slightly less accurate, the size of the linear system matrix resulting from using the coarse mesh ($12354\times 12354$) is remarkably smaller than the ones resulting from the locally refined mesh ($15744\times 15744$) and the single (conforming) mesh ($17415\times 17415$).

We now look into the issue whether the impedance matrices resulting from the MoM discretization of the multiple-scattering EFIE formulation on composite surfaces have spectral properties and condition numbers similar to those resulting from the EFIE formulation on a single surface, when applied to the same problem. In order to address this question, which is relevant for the design of effective preconditioners~\cite{andriulli2008multiplicative,adams2004physical,contopanagos2002well,borel2005new,stephanson2009preconditioned} for the solution of the linear system~\eqref{eq:lin_sym} using accelerated iterative solvers~\cite{song1997multilevel,song1995multilevel,chew2001fast,chew1997fast,seo2005fast}, we present Fig.~\ref{fig:eigsSingle} which displays the eigenvalues of the impedance matrices $Z$, corresponding to the single and composite surface representation of the city-like structure in~Fig.~\ref{fig:city_meshes}.
As can be seen in~Fig.~\ref{fig:eigsSingle}, similar eigenvalue clustering patterns are observed for the two surface representations, with eigenvalues accumulating on the real axis and around  the origin. This suggests that Calder\'on-type preconditioners should in principle improve the convergence of Krylov subspace linear algebra solvers when applied to matrices associated to the multiple-scattering EFIE formulation used on composite surfaces. The matrix condition number in the infinity (resp. one) norm, on the other hand, which amounts to $\kappa_\infty(Z)\approx 362$ (resp.~$\kappa_1(Z)\approx 307$) for the composite surface and  to $\kappa_\infty(Z)\approx 235$ (resp.~$\kappa_1(Z)\approx425$) for the single surface, are similar in this example. However, larger condition numbers are in general obtained for composite surface representations. A natural preconditioner for the multiple-scattering EFIE formulation is the block-diagonal proconditioner resulting from directly inverting the matrix blocks corresponding to the closed surface components of the composite surface. We assess the effectiveness of this preconditioner in this example problem by applying GMRES~\cite{saad1986gmres} directly to the linear system resulting from the EFIE formulation applied on the single-surface, and to the preconditioned system corresponding to the multiple-scattering EFIE formulation applied on the composite surface. For a tolerance of $10^{-4}$, GMRES required 982 iterations in the single-surface case and just 33 iterations in the preconditioned composite-surface case. The eigenvalues of the preconditioned system are shown in Fig~\ref{fig:eigsComp}. The effective and efficient preconditioning of impedance matrices resulting from the multiple-scattering EFIE formulation on composite surfaces is a matter of ongoing research.

\subsubsection{Innovation Center building} 
\begin{figure}[!t]
\centering
\subfloat[]{\includegraphics[scale=0.55]{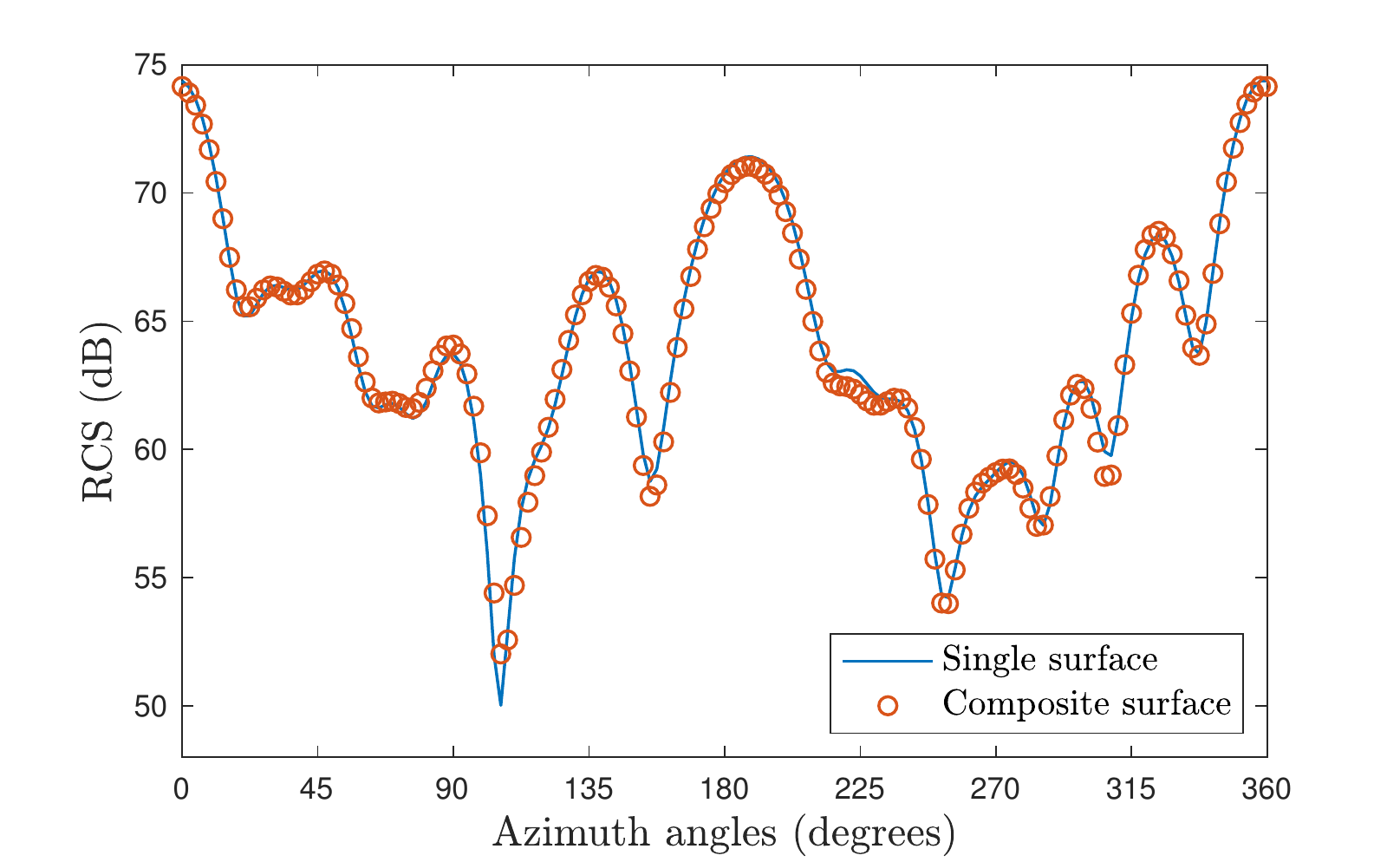}\label{fig:CI_az}}\vspace{-0.2cm}\\
\subfloat[]{\includegraphics[scale=0.55]{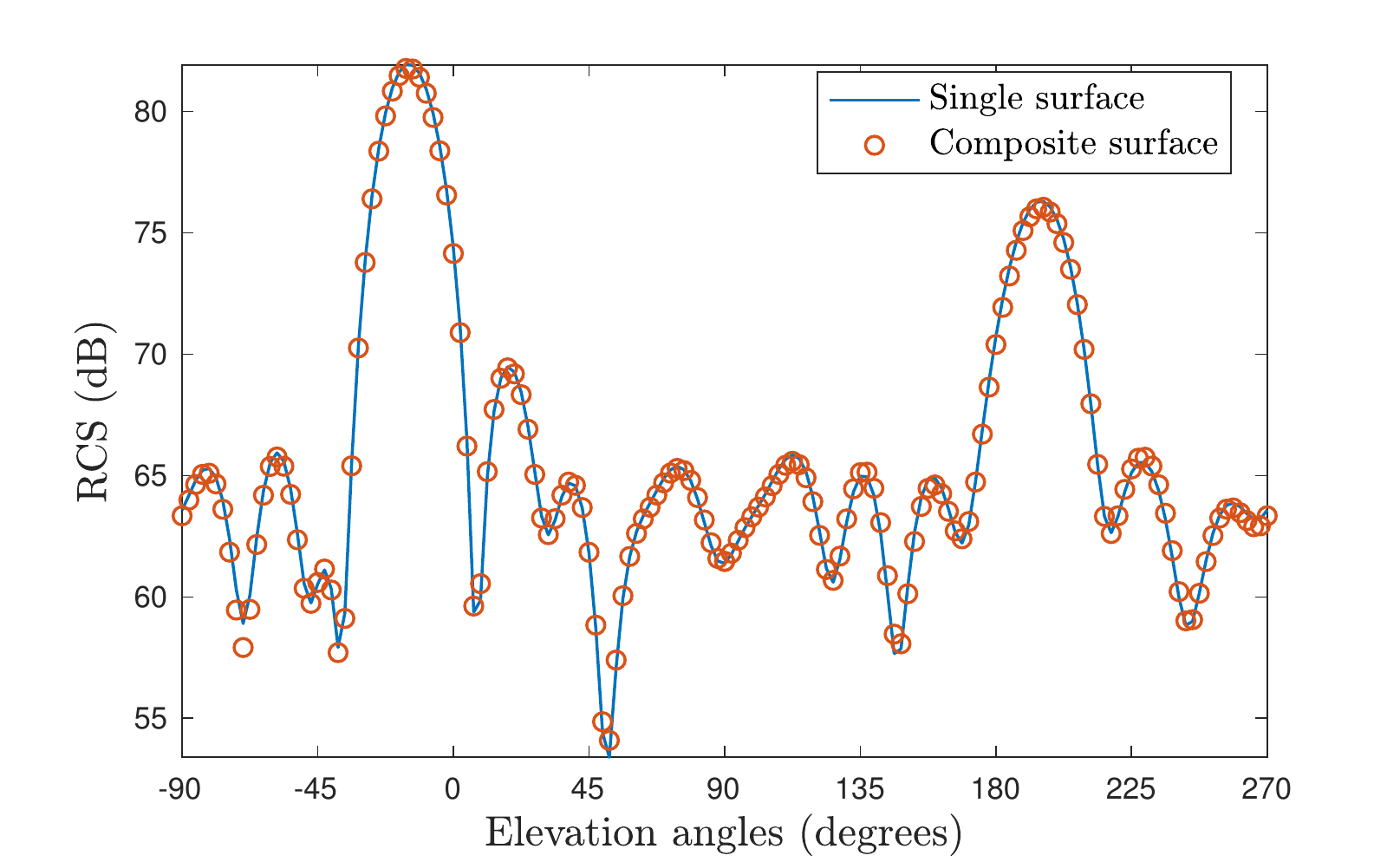}\label{fig:CI_ele}}
\caption{RCSs corresponding to the scattering of a planewave from the building shown in Fig.~\ref{fig:IC}, at zero elevation angle (a), and at zero azimuth angle (b).}
\label{fig:CI_RCS}
\end{figure}

This example considers the intricate geometry of the building shown in Fig.~\ref{fig:IC_photo} whose facade consists of sixteen rectangular blocks of various sizes. The problem of scattering of the planewave $\elf^\inc (\ner)=(\bol p\times \bol d)\e^{ik\bol d \cdot \ner}$, with $f\approx17.8$MHz ($k=\frac{2\pi}{\lambda}$ , $\lambda = 19\,$m), $\bol d=(4,0,-1)/\sqrt{17}$ and $\bol p=(1,1,1)$,  from this structure is here solved by means of both the composite- and the standard-EFIE formulations using the MoM with PWDI kernel-regularized operators and RWG basis functions. The overall size of the composite non-conforming   (resp. single conforming) mesh is $h=2.34$m (resp. $h=2.49$m), which leads to a $16206\times 16206$ (resp. $16203\times 16203$) linear systems matrix. The seemingly complex composite non-conforming mesh shown in Fig.~\ref{fig:IC_meshes} was rendered utilizing Gmsh software~\cite{geuzaine2009gmsh} (http://gmsh.info), which enables the mesh to be created by simply scaling, rotating, and translating a cube. The associated single conforming mesh was produced, on the other hand, by carefully and painstakingly defining the (non-Lipschitz) closed surface of the whole structure, also using Gmsh. Figure~\ref{fig:IC_NF} displays the real part of the three Cartesian components of the total electric field $\elf = \elf^s+\elf^\inc$ on the horizontal plane intersecting the building at 22.5m height, which was evaluated by means of the kernel-regularized off-surface EFIE operator using the high-order algebraic PWDI with orders $M_1=M_2=3$. The accuracy of this near-field calculation can be appraised in detail in the plot of the real part of $E_z$ where it can be seen that the numerically generated $E_z$ remains smooth at and around the surface and, furthermore, it vanishes exactly on the surface, as it is supposed to do in view of the PEC boundary condition~\eqref{eq:scattered_maxwell_3D_BC}. (A similar observation can be made about Fig.~\ref{fig:scatt_sphere}a.)  The resulting bistatic RCSs obtained using the two formulations are shown in Fig.~\ref{fig:CI_az} and~\ref{fig:CI_ele} for the azimuth angles at zero elevation angle, and for the elevation angles at zero azimuth angle, respectively. Excellent agreement of the two solutions is obtained for this challenging geometric setup.

\subsection{Switched parasitic antenna on a finite ground plane} 
\begin{figure}[!t]
\centering
\subfloat[]{\includegraphics[scale=0.25]{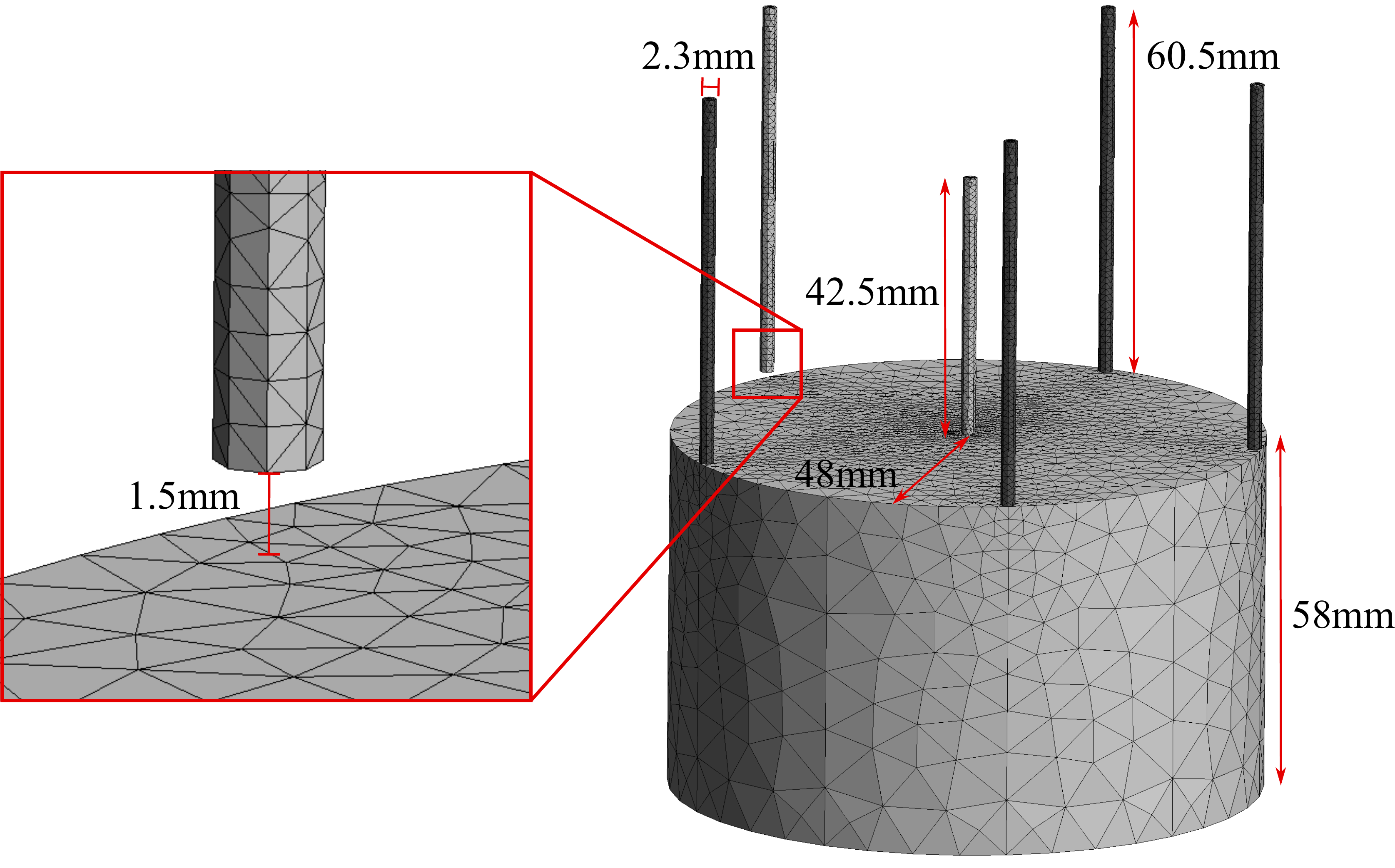}\label{fig:PA}}\\
\subfloat[]{\includegraphics[scale=0.52]{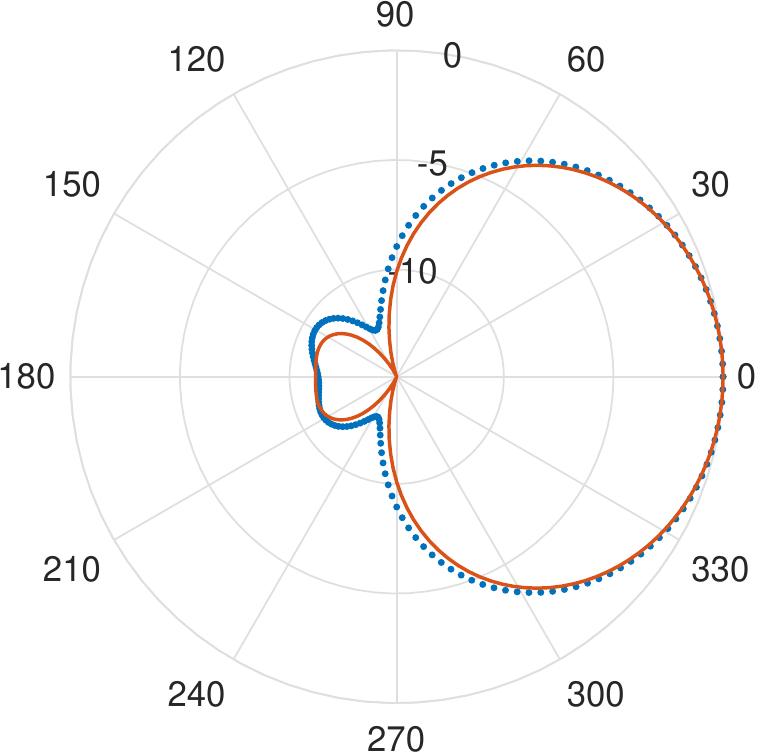}\label{fig:PA_az}}\quad
\subfloat[]{\includegraphics[scale=0.52]{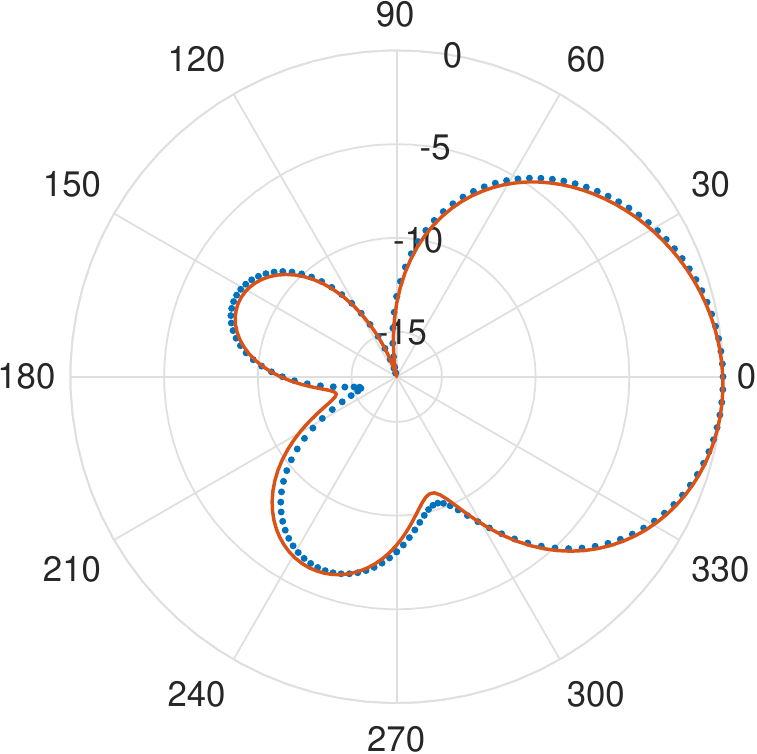}\label{fig:PA_ele}}
\caption{(a) Composite surface mesh of the switched parasitic antenna proposed in~\cite{schlub2004switched}. (b)-(c): Normalized radiation patterns for (b) azimuth and (c) elevation angles computed using the PWDI method (dotted lines). The solid lines correspond to reference radiation patterns computed using significantly refined conforming meshes.}
\label{fig:parasitic_antenna}
\end{figure}

The final example of the paper considers the switched parasitic antenna proposed in~\cite{schlub2004switched} which is shown in~Fig.~\ref{fig:PA}. The antenna is comprised of a small circular finite ground plane with a conductive sleeve, a center monopole feed element attached to the ground, four parasitic reflector elements also attached to the ground and uniformly distributed along its edge, and a fifth parasitic element isolated from the ground plane. The latter corresponds to the switch element that allows steering the radiation along the azimuth in its direction. The precise spatial dimensions of the antenna, which are provided in~\cite[Table~II]{schlub2004switched}, were optimized in order to depress the main lobe elevation. In order to simulate the antenna using the multiple-scattering EFIE formulation~\eqref{eq:E_EFIE_comp}, the delta gap model ~\cite{volakis2012integral} is adopted to represent the monopole feed element. As indicated in~\cite[Fig.~2]{schlub2004switched}, on the other hand, the switch element is realized by simply raising it 1.5mm above the ground plane (see~Fig.~\ref{fig:PA}). The antenna is represented in this example as a composite surface mesh comprising six closed surface meshes corresponding to the skirted finite ground plane with the feed element attached, and the five parasitic cylindrical elements. The actual meshes used in the calculations are shown in Fig.~\ref{fig:PA}. The EFIE is then numerically solved by means of the proposed PWDI technique. Only two meshes are needed to form the impedance matrix $Z$, as by design the five parasitic elements are identical. This allows $Z$ (of size $14793\times 14793$) to be efficiently computed by taking advantage of the fact that the diagonal blocks corresponding to the parasitic elements are also identical in this case. The resulting antenna radiation patterns at the operation frequency 1.575GHz are shown in Figs~\ref{fig:PA_az} and~\ref{fig:PA_ele}, along with reference radiation patterns computed using two significantly refined conforming meshes corresponding the switch element and the rest of the antenna (the impedance matrix associated to the reference solution has size $20103\times 20103$). Note that excellent agreement with both the experimental and the finite-element-produced radiation patterns shown in~\cite[Fig. 4]{schlub2004switched}, is achieved. 

\section{Conclusions}
This paper extended the PWDI methodology put forth in~\cite{plane-wave:2018} and \cite{perez2019planewave} to the classical EFIE formulation for scattering from closed PEC surfaces. We show that the PWDI methodology enables the direct evaluation of Galerkin-MoM impedance matrices using standard quadrature rules, thus significantly simplifying the practical implementation of the MoM. The ability of our method to simultaneously evaluate accurately singular and nearly singular integrals regardless the singularity location allowed us to introduce  a novel EFIE formulation based on non-overlapping subdomain partitioning and use of composite surface representations. This new formulation has the capability of simplifying the geometric treatment of complex three-dimensional structures by enabling the use of non-conforming surface meshes. The advantages of this formulation were demonstrated by applying it to a multiscale structure and an intricate (non-Lipschitz) surface modeling an actual building facade. 

The proposed methodology opens up multiple future research directions. We first mention the immediate extensions/modifications of the PWDI technique to the magnetic field integral equation (MFIE) formulation for PEC scattering problems, and to the Poggio-Miller-Chang-Harrington-Wu-Tsai (PMCHWT)~\cite{poggio1973integral} and M\"uller~\cite{muller2013foundations} formulations for electromagnetic transmission problems.  Yet another research direction has to do with the extension of the proposed PWDI methodology to problems involving unbounded (non-periodic) material interfaces, such as half-spaces, layered media, and waveguides. Current research efforts by the authors in this direction include combining the PWDI technique with the Windowed Green Function method for frequency-~\cite{Bruno2015windowed,bruno2017windowed,bruno2017waveguides} and time-domain~\cite{labarca2019CQ} scattering problems, so as to produce a general-purpose robust and efficient integral equation solver based solely on the free-space Helmholtz Green function.


\appendices


\section*{Acknowledgment}
C.P.-A. gratefully acknowledges support from FONDECYT through Grant No. 11181032. C.T. gratefully acknowledges support from NSF through contracts DMS-1614270 and DMS-1908602. C.S. gratefully acknowledges support by the NSF under Grant No. 1849965.




%
\bibliographystyle{IEEEtran}
\bibliography{IEEEabrv,References}

\end{document}